\documentclass[final]{elsarticle}
\usepackage{lscape}
\usepackage{graphicx}

\usepackage{comment}

\usepackage{zed-csp}

\usepackage{IssamCh3}

\usepackage{mythesis}

\usepackage{datarefine}

\usepackage{setspace}

\usepackage{textcomp}
\usepackage{lineno,hyperref}

\def\pp {{+ \hspace * {-1mm}+ }}
\begin{document}

%%%%%%%%%%%%%%%%%%%%%%%%%%%%%%%%%%%%%%%%%%%%%%%%%%%%%%%%%%%%%%%%%%%%%%%%%%%%%%%%%

\begin{frontmatter}

{\singlespace

\title{The link to the formal publication is via\\ {\small \url{https://doi.org/10.1016/j.advengsoft.2006.01.009}}\\ .\\ Parallel Algorithms Development for Programmable Logic Devices}

    \author{Issam W. Damaj}

    \address{Electrical and Computer Engineering Department, Hariri Canadian Academy of Sciences and
    Technology, Meshref P.O.Box: 10 Damour- Chouf 2010 Lebanon, damajiw@hariricanadian.edu.lb}

\begin{abstract}
Programmable Logic Devices (\textit{PLDs}) continue to grow in size and currently contain several
millions of gates. At the same time, research effort is going into higher-level hardware synthesis
methodologies for reconfigurable computing that can exploit \textit{PLD} technology. In this
paper, we explore the effectiveness and extend one such formal methodology in the design of
massively parallel algorithms. We take a step-wise refinement approach to the development of
correct reconfigurable hardware circuits from formal specifications. A functional programming
notation is used for specifying algorithms and for reasoning about them. The specifications are
realised through the use of a combination of function decomposition strategies, data refinement
techniques, and off-the-shelf refinements based upon higher-order functions. The off-the-shelf
refinements are inspired by the operators of Communicating Sequential Processes (\textit{CSP}) and
map easily to programs in \textit{Handel-C} (a hardware description language). The
\textit{Handel-C} descriptions are directly compiled into reconfigurable hardware. The practical
realisation of this methodology is evidenced by a case studying the matrix multiplication
algorithm as it is relatively simple and well known. In this paper, we obtain several hardware
implementations with different performance characteristics by applying different refinements to
the algorithm. The developed designs are compiled and tested under \textit{Celoxica's}
\textit{RC-1000} reconfigurable computer with its 2 million gates \textit{Virtex-E} \textit{FPGA}.
Performance analysis and evaluation of these implementations are included.
\end{abstract}

}%end singlespace

\end{frontmatter}

%%%%%%%%%%%%%%%%%%%%%%%%%%%%%%%%%%%%%%%%%%%%%%%%%%%%%%%%%%%%%%%%%%%%%%%%%%%%%%%%
\section{Introduction}

The rapid progress and advancement in electronic chips technology
provides a variety of new implementation options for system
engineers. The choice varies between the flexible programs running
on a general purpose processor \textit{(GPP)} and the fixed
hardware implementation using an application specific integrated
circuit (\textit{ASIC}). Many other implementation options
present, for instance, a system with a \textit{RISC} processor and
a \textit{DSP} core. Other options include graphics processors and
microcontrollers. Specialist processors certainly improve
performance over general-purpose ones, but this comes as a quid
pro quo for flexibility. Combining the flexibility of
\textit{GPPs} and the high performance of \textit{ASICs} leads to
the introduction of reconfigurable computing (\textit{RC}) as a
new implementation option with a balance between versatility and
speed.

Generally, reconfigurable computing is computer processing with
highly flexible computing fabrics. The principal difference when
compared to using ordinary microprocessors is the ability to make
substantial changes to the data path itself in addition to the
control flow. In the last decade, there was a renaissance in the
area of reconfigurable computing research with many proposed
reconfigurable architectures developed both in industry and
academia such as, \textit{Matrix}, \textit{Garp}, \textit{RAW},
\textit{DPGA}, \textit{RaPiD}, \textit{PRISM}, \textit{Pleiades},
and \textit{Morphosys} \cite{59}. Such designs were feasible due
to the relentless progress of silicon technology that allowed
complex designs to be implemented on a single chip.

Field Programmable Gate Arrays (\textit{FPGAs}), nowadays are important components of
\textit{RC}-systems, have shown a dramatic increase in their density over the last few years. For
example, companies like \textit{Xilinx} \cite{126} and \textit{Altera} \cite{127} have enabled the
production of \textit{FPGAs} with several millions of gates, such as in \textit{Virtex-II Pro} and
\textit{Stratix-II} \textit{FPGAs}. The versatility of \textit{FPGAs}, opened up completely new
avenues in high-performance computing. These reconfigurable digital electronic hardware circuits
can be combined with high-level software and design methodologies to form a powerful paradigm for
computing.

The traditional implementation of a function on an \textit{FPGA}
is done using logic synthesis based on \textit{VHDL}, Verilog or a
similar \textit{HDL} (hardware description langauge). These
discrete event simulation languages are rather different from
languages, such as \textit{C}, \textit{C++} or \textit{JAVA}. Many
\textit{FPGA} implementation tools are primarily
\textit{HDL}-based and not well integrated with high-level
software tools. Furthermore, these \textit{HDL}-based \textit{IP}
(intellectual property) cores are expensive and they have complex
licensing schemes \cite{128}. These obstacles had caused some
blockage to the infiltration of \textit{FPGAs} as the main
platform solution for hardware engineers. An interesting step
towards more success in hardware compilation is to grant a
higher-level of abstraction from the point of view of programmer.
Designer productivity can be improved and time-to-market can be
reduces by making hardware design more like programming in a
high-level langauge. Recently, vendors have initiated the use of
high-level languages dependent tools like \textit{Handel-C}
\cite{4,129,55,125}, \textit{Forge} \cite{130}, \textit{Nimble}
\cite{131,132}, \textit{SystemC} \cite{133} and \textit{Viva}
\cite{newrev1} (an object-oriented graphical development
environment for programming \textit{FPGAs}).

With the availability of powerful high-level tools accompanying
the emergence of multi-million \textit{FPGA} chips, more emphasis
should be placed on affording an even higher level of abstraction
in programming reconfigurable hardware. Building on these research
motivations, in the work in hand, we extend and examine a
methodology whose main objective is to allow for a higher-level
correct synthesis of massively parallel algorithms and to map
(compile) them onto reconfigurable hardware. Our main concern is
with behavioural refinement, in particular the derivation of
parallel algorithms. The presented methodology systematically
transforms functional specifications of algorithms into parallel
hardware implementations. It builds on the work of Abdallah and
Hawkins \cite{6,9,140,140b} extending their treatment of data and
process refinement.

This paper is divided so that the following section introduces the
adopted development methodology. Section~\ref{bkgnd} presents the
theoretical background. In Section~\ref{App}, we put some emphasis
on the approach to develop different implementations of the matrix
multiplication algorithm. The following section details the
development steps. Section~\ref{RHI} demonstrates selected
implementations. In Section~\ref{PAE}, we analyze and evaluate the
performance of the suggested implementations. Finally,
Section~\ref{Con} concludes the paper.

%%%%%%%%%%%%%%%%%%%%%%%%%%%%%%%%%%%%%%%%%%%%%%%%%%%%%%%%%%%%%%%%%%%%%
\begin{comment}

At this point, many research opportunities present themselves, for
instance, the development methods that allow for the systematic
construction of parallel implementations for various
architectures. One possibility is to parallelise basic building
blocks by instantiating higher-order skeletons. This method is to
realise the generated implementation by automatic mapping onto a
back-end reconfigurable hardware. Another important question could
be raised as to how to exploit the inherent parallelism (software
parallelism) present in computationally intensive applications. In
addition to how to guarantee the correctness of the developed
implementations apart from trial and error. Many more issues could
be questioned, that are of no less significance, like scalability
and reusability of the developed designs.

\end{comment}
%%%%%%%%%%%%%%%%%%%%%%%%%%%%%%%%%%%%%%%%%%%%%%%%%%%%%%%%%%%%%%%%%%%%%

%%%%%%%%%%%%%%%%%%%%%%%%%%%%%%%%%%%%%%%%%%%%%%%%%%%%%%%%%%%%%%%%%%%%%%%%%%%%%%%%

\section{The Development Method}

Although compilers can expose parallelism through data flow
analysis \cite{B4}, imperative languages are perhaps not ideal as
a starting point. This is because imperative programs already
incorporate design decisions (concerning control flows and data
structures), preconditions (that can be assumed), post-conditions
(that must be achieved), and invariants (that must be maintained).
The direct manipulation of state makes it both difficult to prove
that any two pieces of code are equivalent, and to perform
substitutions, modify and rewrite the algorithm. Functional
languages \cite{10}, such as \textit{Haskell} \cite{11}, however,
do not manipulate state directly, and as such gain the property of
referential transparency. Any sub-expression of an algorithm can
be substituted for any other that is provably equivalent. This is
aided by an effective set of laws given to us by such reasoning
frameworks as \textit{Bird-Merteen} Formalism (\textit{BMF})
\cite{Brd87}, along with a wealth of other work in the functional
programming and parallel processing fields
\cite{13,14,15,Rbh95,RnW95,Skl94}.

Although, many hardware development methods still use the powerful
data flow analysis, such as Viva \cite{newrev1}, the attractions
for using the functional paradigm has incited many researchers.
This triggered many investigations in this area, such as
\textit{Lava} \cite{122}, \textit{Hawk} \cite{r1,r2},
\textit{Hydra} \cite{r4}, \textit{HML} \cite{r7}, \textit{MHDL}
\cite{r6}, \textit{DDD} system \cite{r8}, \textit{SAFL}
\cite{r10}, \textit{MuFP} \cite{r11}, \textit{Ruby} \cite{r12},
and \textit{Form} \cite{r14}.

\begin{figure}[h]
	\begin{center}
		\includegraphics [scale=1]%[height=2in,width=2in,angle=0]
		{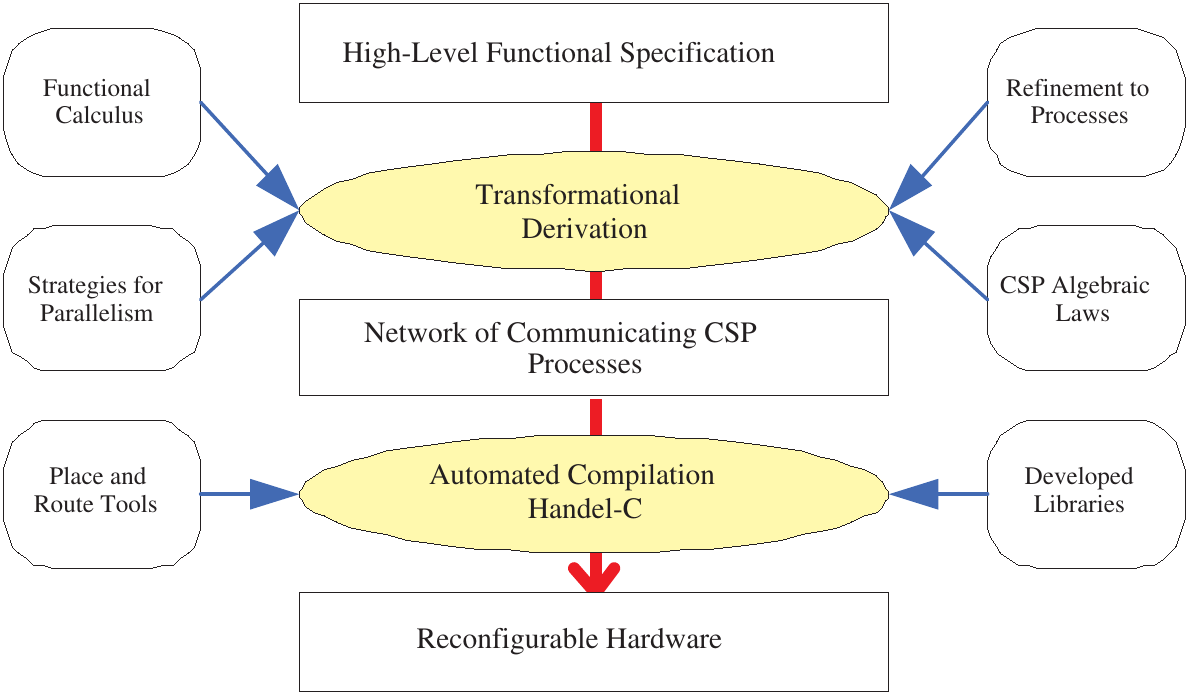}
		\caption{An overview of the transformational derivation and the hardware realisation processes.}
		\label{RDM}
	\end{center}
\end{figure}

The suggested development model adopts the transformational
programming approach for deriving massively parallel algorithms
from functional specifications (See Figure~\ref{RDM}). The
functional notation is used for specifying algorithms and for
reasoning about them. This is usually done by carefully combining
a small number of higher-order functions that serve as the basic
building blocks for writing high-level programs. The systematic
methods for massive parallelisation of algorithms work by
carefully composing an "off-the-shelf" massively parallel
implementation of each of the building blocks involved in the
algorithm. The underlying parallelisation techniques are based on
both pipelining and data parallelism.

Higher-order functions, such as \textit{map}, \textit{filter},
\textit{foldl}, and \textit{foldr}, provide a high degree of
abstraction in functional programs \cite{11}. Not only they do
allow clear and succinct specifications for a large class of
algorithms, but they also are ideal starting points for generating
efficient implementations by a process of mathematical calculation
using \textit{BMF}. Over the past decade, there have been attempts
to apply \textit{BMF} for generating data parallel programs from
abstract specifications using the skeleton approach \cite{15,9}.
The main attraction of this approach is the potential for
increasing reusability of parallel programs without sacrificing
too much performance. The essence of this approach is to design a
generic solution once, and to use instances of the design many
times for various applications. Accordingly, this approach allows
portability by implementing the design on different parallel
architectures.

In order to develop generic solutions for general parallel
architectures, it is necessary to formulate the design within a
concurrency framework such as \textit{CSP} \cite{9,125}. Often
parallel functional programs show peculiar behaviours which are
only understandable in the terms of concurrency rather than
relying on hidden implementation details. The formalisation in
\textit{CSP} (of the parallel behaviour) leads to better
understanding and allows for analysis of performance issues. The
establishment of refinement concepts between functional and
concurrent behaviours may allow systematic generation of parallel
implementations for various architectures. This gives the ability
to exploit well-established functional programming (FP) paradigms
and transformation techniques in order to develop efficient
\textit{CSP} processes. These systematic refinement rules refine
the specification to what we call the \textit{CSP} implementation
stage, where parallelism will be described using Hoare's
\textit{CSP} \cite{16}. Again, this allows issues of immense
practical importance, such as, the careful reasoning about data
distribution, network topology, and locality of communications.

The previous stages of development require a back-end stage for
realising the developed designs. We note at this point that the
\textit{Handel-C} language relies on the parallel constructs in
\textit{CSP} to model concurrent hardware resources. Mostly,
algorithms described with \textit{CSP} could be implemented with
\textit{Handel-C}. Accordingly, this langauge is suggested as the
final reconfigurable hardware realisation stage in the proposed
methodology. It is noted that, for the desired hardware
realisation, \textit{Handel-C} enables the integration with
\textit{VHDL} and \textit{EDIF} (Electronic Design Interchange
Format) and thus various synthesis and place-and-route tools.

%%%%%%%%%%%%%%%%%%%%%%%%%%%%%%%%%%%%%%%%%%%%%%%%%%%%%%%%%%%%%%%%%%%%%%%%%%%%%%%%%

\section{Background}\label{bkgnd}
Abdallah and Hawkins defined in \cite{140} some constructs used in the development model. Their
investigation looked in some depth at data refinement; which is the means of expressing structures
in the specification as communication behaviour in the implementation.

%%%%%%%%%%%%%%%%%%%%%%%%%%%%%%%%%%%%%%%%%%%%%%%%%%%%%%%%%%%%%%%%%%%%%%%%%%%%%%%%%

\subsection{Data Refinement}

In the following we present some datatypes used for refinement, these are stream, vector, and
combined forms.

The stream is a purely sequential method of communicating a group of values. It comprises a
sequence of messages on a channel, with each message representing a value. Values are communicated
one after the other. Assuming the stream is finite, after the last value has been communicated,
the end of transmission (EOT) on a different channel will be signaled. Given some type A, a stream
containing values of type A is denoted as $\langle A \rangle$.

Each item to be communicated by the vector will be dealt with independently in parallel. A vector
refinement of a simple list of items will communicate the entire structure in a single. Given some
type A, a vector of length n, containing values of type A, is denoted as $\lfloor A \rfloor_{n}$.

Whenever dealing with multi-dimensional data structures, for example, lists of lists,
implementation options arise from differing compositions of our primitive data refinements -
streams and vectors. Examples of the combined forms are the Stream of Streams, Streams of Vectors,
Vectors of streams, and Vectors of Vectors. These forms are denoted by: $\langle
S_{1},S_{2},...,S_{n}\rangle$ , $\langle V_{1},V_{2},...,V_{n}\rangle$, $\lfloor
S_{1},S_{2},...,S_{n}\rfloor$, and $\lfloor V_{1},V_{2},...,V_{n}\rfloor$.

%%%%%%%%%%%%%%%%%%%%%%%%%%%%%%%%%%%%%%%%%%%%%%%%%%%%%%%%%%%%%%%%%%%%%%%%%%%%%%%%%

\subsection{Process Refinement}

The refinement of the formally specified functions to processes is the key step towards
understanding possible parallel behaviour of an implementation. In this section, the interest is
in presenting refinements of a subset of functions - some of which are higher-order. A bigger
refined set of these functions is discussed in \cite{9}.

Generally, These highly reusable building blocks can be refined to \textit{CSP} in different ways.
This depends on the setting in which these functions are used (i.e. with streams, vectors etc.),
and leads to implementations with different degrees of parallelism. Note that we don't use
\textit{CSP} in a totally formal way, but we use it in a way that facilitates the
\textit{Handel-C} coding stage later. Recall for the following subsections that values are
communicated through as an \textit{elements} channel, while a single bit is communicated through
another \textit{eotChannel} channel to signal the end of transmission (EOT).

%%%%%%%%%%%%%%%%%%%%%%%%%%%%%%%%%%%%%%%%%%%%%%%%%%%%%%%%%%%%%%%%%%%%%%%%%%%%%%%%%

\subsubsection{Produce}

The produce process (\textit{PRD}) is fundamental to process refinement. It is used to produce
values on the channels of a certain communication construct (\textit{Item}, \textit{Stream},
\textit{Vector}, and so on). These values are to be received and manipulated by another processes.

%%%%%%%%%%%%%%%%%%%%%%%%%%%%%%%%%%%%%%%%%%%%%%%%%%%%%%%%%%%%%%%%%%%%%%%%%%%%%%%%%

\paragraph{Items}

\begin{figure}[htpb]
	\begin{center}
		\includegraphics [scale=1]%[height=2in,width=2in,angle=0]
		{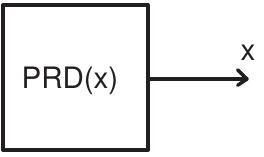}
		\caption{The Produce process (PRD) for items}
		\label{PRDx}
	\end{center}
\end{figure}

For simple, single item types (\textit{int}, \textit{char}, \textit{bool}, etc.), the produce
process is very simple. This is depicted in Figure~\ref{PRDx}. Here the output is just a single
channel.

The definition in \textit{CSP} notation is very straightforward:

\[
PRD ~ (Item ~ a) = out.element.channel ~ ! ~ a \rightarrow SKIP
\]

%%%%%%%%%%%%%%%%%%%%%%%%%%%%%%%%%%%%%%%%%%%%%%%%%%%%%%%%%%%%%%%%%%%%%%%%%%%%%%%%%

\paragraph{Streams}

\begin{figure}[htpb]
	\begin{center}
		\includegraphics [scale=1]%[height=2in,width=2in,angle=0]
		{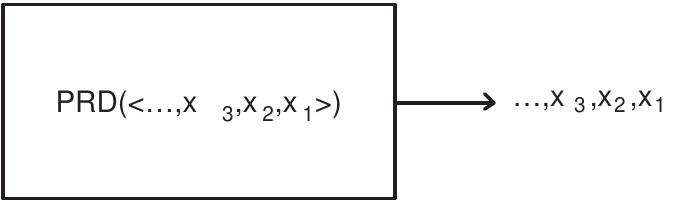}
		\caption{The Produce process (PRD) for streams}
		\label{PRDs}
	\end{center}
\end{figure}

The produce process for streams is depicted in Figure~\ref{PRDs}. As already noted, the output in
this case is a pair of two other channels. One channel will produce the values of the stream, and
the other will be a simple channel used to signal \textit{EOT}.

In a more general case, the structure of the values which the stream is carrying is not
necessarily known. These may be simple items, but may also be streams or vectors. Generally,
producing a stream could be described as:

\begin{equ}
PRD ~ (\stream{s}) = ((;)_{i=1}^{i=length(s)} (PRD ~ s_{i}) [out.elements.channel / out]);

\\
out.eotChannel ~ ! ~ eot \rightarrow SKIP
\end{equ}

%%%%%%%%%%%%%%%%%%%%%%%%%%%%%%%%%%%%%%%%%%%%%%%%%%%%%%%%%%%%%%%%%%%%%%%%%%%%%%%%%

\paragraph{Vectors}

For vectors of size $n$, $n$ instances of the produce process are composed in parallel, one for
each item in the vector. The output here is an array of channels. This is depicted in
Figure~\ref{PRDv}. A general definition is given below:

\begin{equ}
PRD ~ (\lfloor v \rfloor_{n}) & = & \interleave_{i=1}^{i=n} (PRD ~ v_{i}) [
out.elements_{i}.channel / out ]
\end{equ}

A process \textit{STORE} stores a communication construct in a variable. We use this process to
store items, vectors, streams, or combinations. A subscript letter is used with the processes
\textit{PRD} and \textit{STORE} to indicate the type of communication. We sometimes omit this
subscript if the communication structure is clear from context.

\begin{figure}[htpb]
	\begin{center}
		\includegraphics [scale=1]%[height=2in,width=2in,angle=0]
		{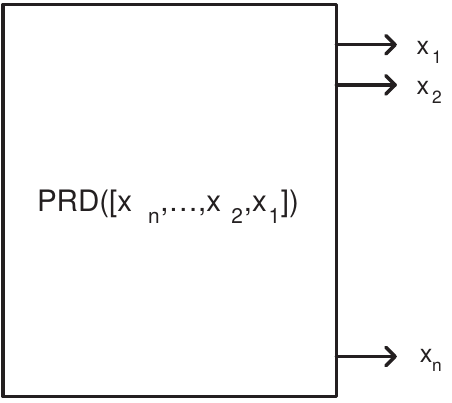}
		\caption{The Produce process (PRD) for vectors}
		\label{PRDv}
	\end{center}
\end{figure}

%%%%%%%%%%%%%%%%%%%%%%%%%%%%%%%%%%%%%%%%%%%%%%%%%%%%%%%%%%%%%%%%%%%%%%%%%%%%%%%%%

\subsubsection{Feeding Processes}

The feed operator in \textit{CSP} models function application. The feed operator is written
$\rhd$. The feed operator takes two processes, composes them together in parallel, and renames
both the output of the first and the input of the second to a new name, which is then hidden.
Given the lifted concepts of \textit{CSP} channel renaming and hiding, the definition can remain
the same regardless of the type of the communicating construct (\textit{Item}, \textit{Stream},
\textit{Vector} or any combination).

\[
P \rhd Q = (P [mid / out] ~ || ~ Q [mid / in]) \verb"\" \{ mid \}
\]

%%%%%%%%%%%%%%%%%%%%%%%%%%%%%%%%%%%%%%%%%%%%%%%%%%%%%%%%%%%%%%%%%%%%%%%%%%%%%%%%%

\subsubsection{Formal Process Refinement}

Given the definition of a feed operator that operates on processes, a formal definition of process
refinement could be delivered. Consider a function $f$, which takes in values of type $A$ and
returns values of type $B$. Assume that the data refinement step has already been performed, such
that $A$ and $B$ are both types of some transmission value:

\[
f :: A \rightarrow B
\]

Then, consider a potential refinement for $f$, a process $F$. The operator $\sqsubseteq$ denotes a
process refinement, where the left hand side is a function, and the right hand side is a process.
To state that $f$ is refined to $F$, or in other words, the process $F$ is a valid refinement of
the function $f$, the following may be used:

\[
f \sqsubseteq F
\]

The rules of refinement were proven once \cite{9}, and in this paper we use them systematically to
refine the functional specification into a network of communicating processes.

%%%%%%%%%%%%%%%%%%%%%%%%%%%%%%%%%%%%%%%%%%%%%%%%%%%%%%%%%%%%%%%%%%%%%%%%%%%%%%%%%

\subsubsection{MAP the Process Refinement of the Higher-order Function \textit{map}}

Now the attention is turned to the refinement of the widely used higher-order function
\textit{map} \cite{140} . Employing this function in stream and vector settings is presented. The
refinement for combined structures is to be made in a similar way.

%%%%%%%%%%%%%%%%%%%%%%%%%%%%%%%%%%%%%%%%%%%%%%%%%%%%%%%%%%%%%%%%%%%%%%%%%%%%%%%%%

\paragraph{Streams}

\begin{figure}[htpb]
	\begin{center}
		\includegraphics [scale=1]%[height=2in,width=2in,angle=0]
		{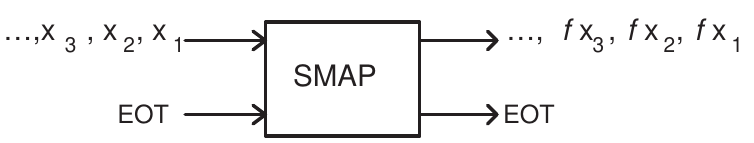}
		\caption{The SMAP process for streams}
		\label{SMAP}
	\end{center}
\end{figure}

A process implementing the functionality of $map ~ f$ in stream terms should input a stream of
values, and output a stream of values with the function $f$ applied (See Figure~\ref{SMAP}).

In general, the handling of the \textit{EOT} channels will be the same. However, the handling of
the value will vary depending on the type of the elements of the input and output stream.

\[
SMAP(F) =\\
\mu  X ~ \bullet  in.eotChannel ~ ? ~ eot \rightarrow out.eotChannel ~ ! ~ eot \rightarrow SKIP \\
                   \choice \\
                   F [in.elements.channel / in, out.elements.channel / out] ; X
\]

%%%%%%%%%%%%%%%%%%%%%%%%%%%%%%%%%%%%%%%%%%%%%%%%%%%%%%%%%%%%%%%%%%%%%%%%%%%%%%%%%

\paragraph{Vectors}

\begin{figure}[htpb]
	\begin{center}
		\includegraphics [scale = 1]
		{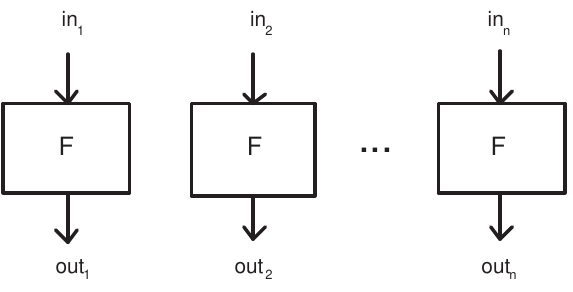}
		\caption{The VMAP process for vectors}
		\label{VMap}
	\end{center}
\end{figure}

In functional terms, the functionality of $map ~ f$ in a list setting is modelled by $vmap ~ f$ in
the vector setting. Consider $F$ as a valid refinement of the function $f$. The implementation of
$VMAP$ can then proceed by composing $n$ instances of $F$ in parallel, and directing an item from
the input vector to each instance for processing (See Figure~\ref{VMap}). In \textit{CSP} we have:

\begin{equ}
VMAP_{n}(F) & = & \interleave_{i=1}^{i=n} F [ in_{i} / in, out_{i} / out]
\end{equ}

%%%%%%%%%%%%%%%%%%%%%%%%%%%%%%%%%%%%%%%%%%%%%%%%%%%%%%%%%%%%%%%%%%%%%%%%%%%%%%%%%

\subsubsection{ZIPWITH the Process Refinement of the Higher-order Function \textit{zipWith}}

Recall another higher-order function, namely \textit{zipWith}. This function is used to zip two
lists (taking one element from each list) with a certain operation. Formally:

\[
zipWith :: (A\rightarrow B\rightarrow C) \rightarrow [A]\rightarrow[B]\rightarrow[C]
\]
\[
zipWith ~ (\oplus) ~ [x_{1}, x_{2}, ... x_{n}] [y_{1}, y_{2}, ... y_{n}] = [x_{1} \oplus y_{1},
x_{2} \oplus y_{2},..., x_{n} \oplus y_{n}]
\]

%%%%%%%%%%%%%%%%%%%%%%%%%%%%%%%%%%%%%%%%%%%%%%%%%%%%%%%%%%%%%%%%%%%%%%%%%%%%%%%%%

\subsubsection{Streams}
\begin{figure}[htpb]
	\begin{center}
		\includegraphics [scale =1]
		{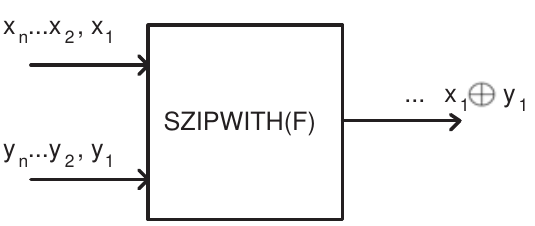}
		\caption{The SZIPWITH process for streams}
		\label{SZipWith}
	\end{center}
\end{figure}

The process implementation of ($zipWith ~ f)$ in stream terms should input two streams of values,
and output a stream of values with the function $f$ applied (See Figure~\ref{SZipWith}).

Again, the handling of the \textit{EOT} channel will be the same. Nevertheless, the handling of
the value will vary depending on the type of the input and output streams elements.

\[
\begin{array}{lcl}
SZIPWITH(F)  =  \\
\mu  X ~ \bullet  in.eotChannel ~ ? ~ eot \rightarrow out.eotChannel ~ ! ~ eot \rightarrow SKIP \\
\choice \\
F [in_{1}.elements.channel / in_{1}, in_{2}.elements.channel /
in_{2}, \\
out.elements.channel / out] ; X
\end{array}
\]

%%%%%%%%%%%%%%%%%%%%%%%%%%%%%%%%%%%%%%%%%%%%%%%%%%%%%%%%%%%%%%%%%%%%%%%%%%%%%%%%%

\subsubsection{Vectors}

\begin{figure}[htpb]
	\begin{center}
		\includegraphics [scale = 1]
		{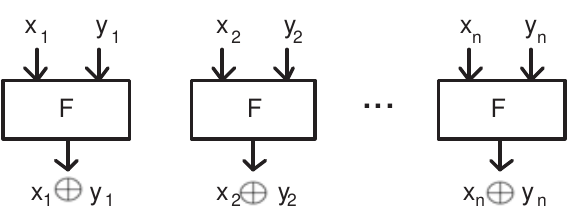}
		\caption{The VZIPWITH process for vectors}
		\label{VZipWith}
	\end{center}
\end{figure}

To implement the data parallel version of this higher-order function, we refine it to a process
\textit{VZIPWITH} that takes two vectors as input and zips the two lists with a process
\textit{F}; \textit{F} is a refined process from the function ($\oplus$). This is depicted as in
Figure~\ref{VZipWith}.

\[
vzipWith ~ (\oplus):: \lfloor A \rfloor_{n} \rightarrow ,\lfloor B \rfloor_{n} \rightarrow \lfloor
C \rfloor_{n}
\]

\[
VZIPWITH ~ (\oplus) = \interleave_{i=1}^{i=n}F[out_{i}/out,c_{i}/in_{1},d_{i}/in_{2}]
\]

%%%%%%%%%%%%%%%%%%%%%%%%%%%%%%%%%%%%%%%%%%%%%%%%%%%%%%%%%%%%%%%%%%%%%%%%%%%%%%%%%

\subsection{Decomposition of Higher-Order Functions for Parallel Programs Derivation} \label{Decmpstn}

Intrinsic richness and usefulness of higher-order functions could be made clear by recalling some
of previous work presented in \cite{8}. This work concentrated on providing systematic
decomposition methods for exploiting pipelined parallelism in instances of the higher-order
function \textit{foldr}. In this section, the decomposition for the higher-order function
\textit{map} is shown. This decomposition rule is used in the forthcoming applications.

\begin{figure}[htpb]
	\begin{center}
		\includegraphics [scale = 1]
		{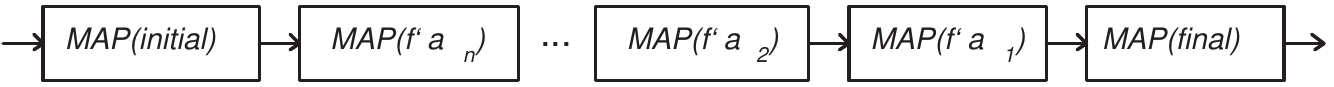}
		\caption{The decomposed network SPEC}
		\label{SPEC}
	\end{center}
\end{figure}

The following decomposition rule is recalled along with its corresponding \textit{CSP}
implementation. This rule decomposes a specification of the form \textit{(map (h m))}, where h is
a function and m is a given list of values. The \textit{CSP} network \textit{SPEC} which refines
this specification is shown in Figure~\ref{SPEC}.
\\

 $spec::A \rightarrow B; h :: [T] \rightarrow A \rightarrow B ;m::[T];e::B$
 $f::T \rightarrow A \rightarrow B \rightarrow B$
 $spec = map$ \ $(h$ \  $m)$
 $h$ \ $[]$ \ $x$ \ $=$ \ $e$
 $h$ \ $(a$ \ $:$ \ $s)$ \  $x$ \ $=$ \ $f$ \ $a$ \ $x$ \ $(h$ \ $s$ \ $x)$

This could be decomposed to:

 $spec = (\circ) / [final^{_{\ast}}]$ \ $\pp$ \ $(map \circ  f' \ast m)$ \ $\pp$ \
 $[initial^{_{\ast}}]$
 $f'$ \ $a$ \ $\langle x , y \rangle$ \ $=$ \ $\langle x,$ \ $f$ \ $a$ \ $x$ \ $y \rangle$
 $initial$ \ $x$ $= \langle x,$ \ $y \rangle$
 $final$ \  $\langle x,$ \ $y \rangle = y$

The pipelined network of \textit{CSP} processes \textit{SPEC}, which refines the functional
specification \textit{spec} is synthesised as follows:

 $SPEC = (\gg)/[MAP(initial)] \concat ((MAP \circ f') \ast (reverse$ \ $m)) \concat [MAP(final)] $
 $MAP(initial)= \mu X$ \ $\bullet$ \ \ \ $left?eot \rightarrow right!eot \rightarrow SKIP$
 \\
 $|$
 \\
 $left?x \rightarrow right! \langle x, e \rangle \rightarrow X$

 $MAP(f'$ \ $as) = \mu X$ \ $\bullet$ \ \ $left?eot \rightarrow right!eot \rightarrow SKIP$
 \\
 $|$
 \\
 $left?\langle x, y \rangle \rightarrow right! \langle x, f$ \ $a$ \ $x$ \ $y \rangle \rightarrow X$

 $MAP(final)= \mu X$ \ $\bullet$ \ \ \ $left?eot \rightarrow right!eot \rightarrow SKIP$
 \\
 $|$
 \\
 $left? \langle x, y \rangle
 \rightarrow right!y \rightarrow X$

It is important not jump to the conclusion that every parallel
algorithm can be deve1oped this way. There are two limitations to
this approach. First, it only deals with systems which can be
specified functionally. Second, it may not be possible to develop
some parallel algorithms which use multi-directional
communications using this method. This second point will be
practically assessed in later section while designing a
multilevel-pipelined parallel program.

%%%%%%%%%%%%%%%%%%%%%%%%%%%%%%%%%%%%%%%%%%%%%%%%%%%%%%%%%%%%%%%%%%%%%%%%%%%%%%%%%

\subsection{Handel-C as a Stage in the Development Model}

Based on datatype refinement and the skeleton afforded by process refinement, the desired
reconfigurable circuits are built. Circuit realisation is done using \textit{Handel-C}, as it is
based on the theories of \textit{CSP} \cite{16} and \textit{Occam} \cite{B3}.

\begin{figure}[htpb]
	\begin{center}
		\includegraphics [scale = 1]
		{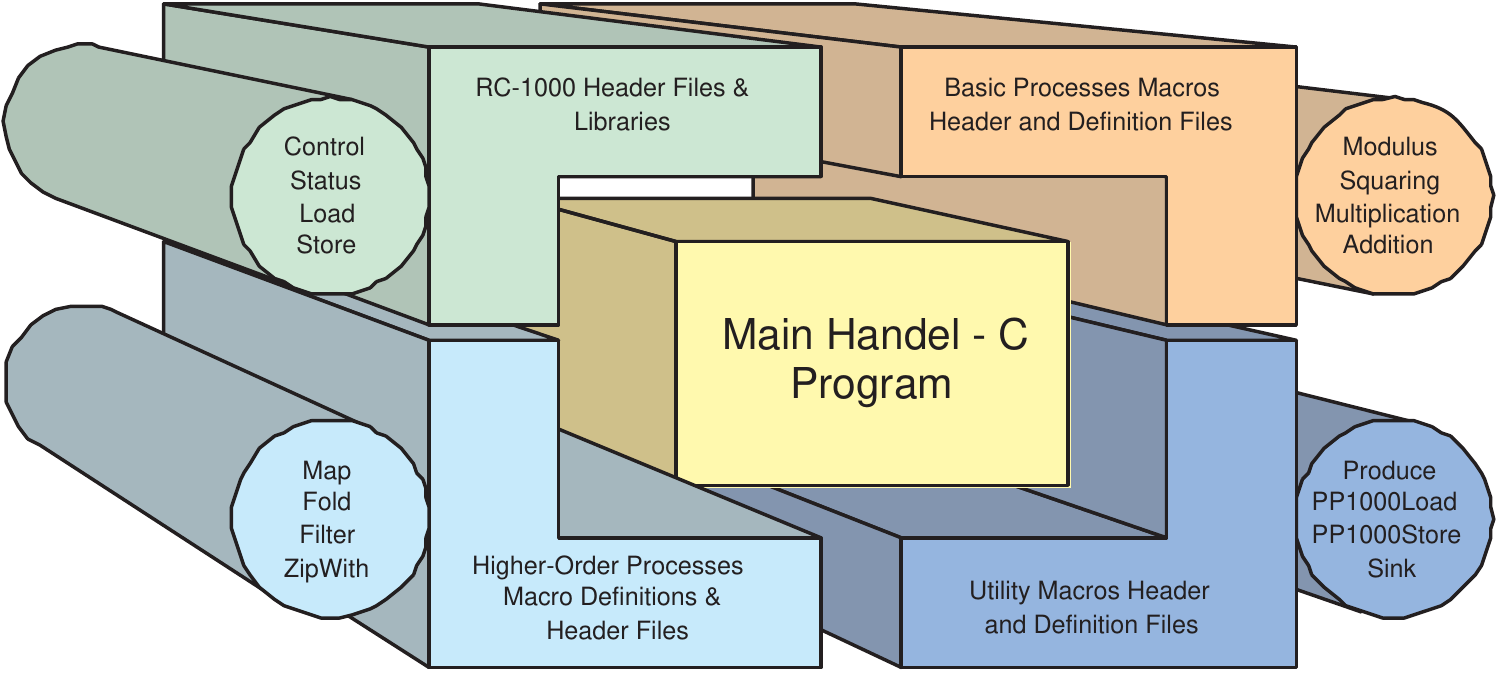}
		\caption{Handel-C code constructs organisation}
		\label{fig14}
	\end{center}
\end{figure}

From a practical standpoint, each refined datatype is defined as a structure in \textit{Handel-C},
while each process is implemented as a \textit{macro} \textit{procedure}. The constructs
corresponding to the \textit{CSP} stage are divided into 2 main categories for organisation
purposes. The first category represents the definitions of the refined datatypes. The second
category implements the refined processes. The refined processes are divided into different
groups. The \textit{utility} \textit{processes} group contains macros responsible for producing,
storing, sinking, broadcasting data and etc. The \textit{basic} \textit{processes} group contains
macros that correspond to simple arithmetic and logical operations. These basic processes could be
simple addition, multiplication, etc. The \textit{higher-order} \textit{processes} group contains
the macros realising the \textit{CSP} implementations corresponding to the higher-order functions.
A separate group contains the macros that handle the \textit{FPGA} card setup and general
functionality. The reusable macros found in these groups serves as building blocks used for
constructing a certain specified algorithm. This organisation is visualised in Figure~\ref{fig14}.

%%%%%%%%%%%%%%%%%%%%%%%%%%%%%%%%%%%%%%%%%%%%%%%%%%%%%%%%%%%%%%%%%%%%%%%%%%%%%%%%%

\subsubsection{Datatypes Definitions}
The datatypes definitions are implemented using structures. This method supports recursive as well
as simple types. The definition for an \textit{Item} of a type \textit{Msgtype} is a structure
that contains a communicating channel of that type.

{\singlespace  {\small
\begin{verbatim}
 #define Item(Name, Msgtype)
    struct {
        chan Msgtype    channel;
        Msgtype         message;
        } Name
\end{verbatim}}}

For generality in implementing processes the type of the communicating structure is to be
determined at compile time. This is done using the \textit{typeof} type operator, which allows the
type of an object to be determined at compile time. For this reason, in each structure we declare
a \textit{message} variable of type \textit{Msgtype}.
\\

A stream of items, called \textit{StreamOfItems}, is a structure with three declarations a
communicating channel, an \textit{EOT} channel, and a \textit{message} variable \cite{140}:

{\singlespace  {\small
\begin{verbatim}
 #define StreamOfItems(Name, Msgtype)
    struct {
        Msgtype         message;
        chan Msgtype    channel;
        chan Bool       eotChannel;
        } Name
\end{verbatim}}}

A vector of items, called \textit{VectorOfItems}, is a structure with a variable \textit{message}
and another array of sub-structure elements \cite{140}.

{\singlespace  {\small
\begin{verbatim}
 #define VectorOfItems(Name, n, Msgtype)
    struct {
        struct {
            chan Msgtype    channel;
            } elements[n];
        Msgtype     message;
        } Name
\end{verbatim}}}

Other definitions are possible, but it affects the way a channel is called using the structure
member operator (.). Examples of different extended definitions are as follows (the first
definition reuses the \textit{Item} structure, while the second one employs channel arrays
supported in \textit{Handel-C}):

{\singlespace  {\small
\begin{verbatim}
 #define VectorOfItems(Name, n, Msgtype)
    struct {
        struct {
            Item(element, MsgType);
            } elements[n];
        } Name

 #define VectorOfItems(Name, n, Msgtype)
    struct {
        chan Msgtype    channel[n];
        Msgtype         messages;
        } Name
\end{verbatim}}}

%%%%%%%%%%%%%%%%%%%%%%%%%%%%%%%%%%%%%%%%%%%%%%%%%%%%%%%%%%%%%%%%%%%%%%%%
\begin{comment}

Samples of some combined structures devised for the use in this
paper are as follows:

{\singlespace  {\small
\begin{verbatim}
 #define StreamOfStreamsOfItems(Name, Msgtype)
    struct {
        struct {
            chan Msgtype    channel;
            chan Bool       eotChannel;
            } elements;
        chan Bool       eotChannel;
        Msgtype     message;
        } Name


 #define StreamOfVectorsOfItems(Name, n, Msgtype)
    struct {
        struct {
            Msgtype         message;
            chan Msgtype    channel;
            } elements [n];
        chan Bool       eotChannel;
        } Name
\end{verbatim}}}

\end{comment}
%%%%%%%%%%%%%%%%%%%%%%%%%%%%%%%%%%%%%%%%%%%%%%%%%%%%%%%%%%%%%%%%%%%%%%%%

In general, there are certain limitations in the \textit{Handel-C}
language, which make the expression of a number of useful
constructs either difficult or impossible. An example of an
impossible to implement declaration is as follows:

\begin{verbatim}
 StreamOfItems(Name, VectorOfItems(Name, n, Int16));
\end{verbatim}

A simple preprocessor would facilitate a higher level of \textit{Handel-C} generic definitions,
and allow them to flow much more freely from our functional specifications. The implementation of
such a preprocessor is being investigated within our research group.

%%%%%%%%%%%%%%%%%%%%%%%%%%%%%%%%%%%%%%%%%%%%%%%%%%%%%%%%%%%%%%%%%%%%%%%%

\subsubsection{Utilities Macros}
The utility processes used in the implementation are related to
the employed datatypes. The \textit{Handel-C} implementation of these processes relies on their
corresponding \textit{CSP} implementation. In the following, we present an instance of these
utility macros.

{\singlespace  {\small
\begin{verbatim}
 macro proc ProduceItem(Item, x){
    Item.channel ! x;}

 macro proc StoreItem(Item, x){
    Item.channel ? x;}
\end{verbatim}}}

%%%%%%%%%%%%%%%%%%%%%%%%%%%%%%%%%%%%%%%%%%%%%%%%%%%%%%%%%%%%%%%%%%%%%%%%
\begin{comment}

In the case of a vector of items, the values are produced (stored) from (in) an array structure in
parallel using the \textit{par} operator through the available communicating channels.

The code that implements these utility macros is as follows:

{\singlespace  {\small
\begin{verbatim}
 macro proc ProduceVectorOfItems (v, n, xs){
    par (i = 0; i < n; i++){
        v.elements[i].channel ! xs[i];}}

 macro proc StoreVectorOfItems(v, n, xs){
    par(i = 0; i < n; i++){
        v.elements[i].channel ? xs[i];}}
\end{verbatim}}}

Items in a stream of items are produced or stored one by one as follows:

{\singlespace  {\small
\begin{verbatim}
 macro proc ProduceStreamOfItems(s, n, xs){
    typeof (n) i;
    for (i = 0; (i < n) ; i++){
        s.channel ! xs[i];
        if (i == n - 1){break;}}
    s.eotchannel ! True;}

 macro proc StoreStreamOfItems(s, n, xs){
    typeof (n - 1) i;
    typeof (s.message) temp;
    Bool eot;
    eot = False;
    i = 0;
    do{
        prialt{
        case s.channel ? temp:
             xs[i] = temp;
             i++;
             break;
        case s.eotchannel ? eot:
             break;}} while (!eot);}
\end{verbatim}}}

%%%%%%%%%%%%%%%%%%%%%%%%%%%%%%%%%%%%%%%%%%%%%%%%%%%%%%%%%%%%%%%%%%%%%%%%
\end{comment}

\subsubsection{Basic Processes Macros}
This group of macros represents the fine-grained processes.
A sample basic macro procedure \textit{Addition} is included as an example.

%%%%%%%%%%%%%%%%%%%%%%%%%%%%%%%%%%%%%%%%%%%%%%%%%%%%%%%%%%%%%%%%%%%%%%%%

{\singlespace  {\small
\begin{verbatim}
 macro proc Addition(xItem, yItem, output){
    typeof (xItem.message) x,y;
    xItem.channel ? x;
    yItem.channel ? y;
    output.channel ! (x + y);}
\end{verbatim}}}

%%%%%%%%%%%%%%%%%%%%%%%%%%%%%%%%%%%%%%%%%%%%%%%%%%%%%%%%%%%%%%%%%%%%%%%%

\subsubsection{Higher-Order Processes Macros}

An example for an implementation in \textit{Handel-C} of the CSP refinement of a higher-order
function (\textit{map}) is done as follows. The process hinges around a loop which terminates when
the variable \textit{eot} is set to true. At each step of the loop, the process enters a wait
state until either the \textit{EOT} or the value channel of the input stream is willing to
communicate. If the \textit{EOT} channel is willing to communicate, the input is consumed from it
and stored in the variable \textit{eot}, then output an \textit{EOT} message for the output
stream.  If the value channel of the input stream is willing to communicate, the value is consumed
then $F$ is applied to it giving the result on the output stream channel \cite{140}.

{\singlespace  {\small
\begin{verbatim}
 macro proc SMAP (streamin, streamout, F){
   Bool eot;
   eot  = False;
   do{
      prialt{
      case streamin.eotChannel ? eot:
         streamout.eotChannel ! True;
         break;
      default:
         F (streamin.elements,streamout.elements);
         break;
      }} while (!eot)}
\end{verbatim}}}

We turn the attention to providing a definition in \textit{Handel-C} for the behaviour of the
process \textit{VMAP}. Here we can employ \textit{Handel-C}'s enumerated \textit{par} construct to
place $n$ instances of the process $F$ in parallel. Each instance is passed to the corresponding
channels from both the input and output conduits \cite{140}.

{\singlespace  {\small
\begin{verbatim}
 macro proc VMAP (n, vectorin, vectorout, F) {
   typeof (n) c;
   par (c = 0 ; c < n ; c++){
      F(vectorin.elements[c], vectorout.elements[c]);}}
\end{verbatim}}}

The implementations of the stream and vector settings of the
remaining high-order functions is done in a similar manner.

%%%%%%%%%%%%%%%%%%%%%%%%%%%%%%%%%%%%%%%%%%%%%%%%%%%%%%%%%%%%%%%%%%%%%%%%
\begin{comment}

 \textit{SZipWith} and \textit{VZipWith} are
straightforward. The process $F$ is said to have two inputs and
one output. The implementation is as follows:

{\singlespace  {\small
\begin{verbatim}
 macro proc SZipWith(s1 , s2, sOutput, F) {
    typeof(s1.message) X;
    Bool eot1;
    Bool eot2;
    eot1 = False;
    eot2 = False;
    do{
        prialt{
        case s1.eotChannel ? eot1:
            s2.eotChannel ? eot2;
            sOutput.eotChannel ! True;
            break;
        default:
            F(s1, s2, sOutput);
            break;}}while (!eot1);}
\end{verbatim}}}

{\singlespace  {\small
\begin{verbatim}
 macro proc VZipWith(n, v1,  v2, vOut, F){
    par(i = 0; i < n; i++){
        F(v1.elements[i], v2.elements[i], v2.elements[i]);}}
\end{verbatim}}}

\end{comment}
%%%%%%%%%%%%%%%%%%%%%%%%%%%%%%%%%%%%%%%%%%%%%%%%%%%%%%%%%%%%%%%%%%%%%%%%

\subsubsection{The RC-1000 System Control Macros}

The \textit{Celoxica} \textit{RC-x000} boards provide
high-performance, real-time processing capabilities and are
optimised for the \textit{Celoxica} \textit{DK} design suite. The
\textit{RC-1000} is a standard \textit{PCI} bus card, with four
onboard banks of SRAM, equipped with a \textit{Xilinx}
\textit{Virtex} with up to 2 million system gates\cite{129}.

According to the characteristics of the used system, some reusable macro procedures were
implemented to be employed in the development model. For instance, reading or storing an
\textit{Item} from (in) a bank could be implemented as in the following macros:

{\singlespace  {\small
\begin{verbatim}
 macro proc ReadItemFromBank1(r){
    Int temp;

    PP1000ReadBank1(0, temp);
    r.channel ! temp;}
\end{verbatim}}}

{\singlespace  {\small
\begin{verbatim}
 macro proc StoreItemToBank1(r) {
    Int temp;
    unsigned int 21 count;

    r.channel? temp;
    PP1000WriteBank1(0, temp);}
\end{verbatim}}}

%%%%%%%%%%%%%%%%%%%%%%%%%%%%%%%%%%%%%%%%%%%%%%%%%%%%%%%%%%%%%%%%%%%%%%%%
\begin{comment}

Technically, accessing an available memory bank is done sequentially. Thus, only vector of Items
with a size of 4 could be produced in parallel directly from the available 4 memory banks. Thus,
normally to produce a vector of items or to store it in memory, an intermediate storage array is
used.

\end{comment}
%%%%%%%%%%%%%%%%%%%%%%%%%%%%%%%%%%%%%%%%%%%%%%%%%%%%%%%%%%%%%%%%%%%%%%%%

\subsection{Evaluation Tools and Performance Metrics}

Different tools are used to measure the performance metrics used for the analysis. These tools
include the design suite (\textit{DK}) from \textit{Celoxica}, where we get the number of
\textit{NAND} gates for the design as compiled to the Electronic Design Interchange Format
(\textit{EDIF}). The \textit{DK} also affords the number of cycles taken by a design using its
simulator. Accordingly, the speed of a design could be calculated depending on the expected
maximum frequency of the design. The maximum frequency could be determined by the timing analyzer.
Accordingly, the time to execute 1 cycle could be determined (\textit{Period}). The execution time
of a design is then the \textit{Period} multiplied by the number of cycles. Thereat, the
throughput is calculated depending on the amount of data processed in that execution time.

To get the practical execution time as observed from the host computer, the \textit{C++}
high-precision performance counter is used. The counter probes the execution of the design after
loading the image of the design into the \textit{FPGA} till termination.

%%%%%%%%%%%%%%%%%%%%%%%%%%%%%%%%%%%%%%%%%%%%%%%%%%%%%%%%%%%%%%%%%%%%%%%%%%%%%%%%%%%
\begin{comment}

Practically, the probation comes directly after writing a control
signal to the \textit{FPGA} enabling execution. The counter stops
immediately after receiving a signal through reading the status
register.

According to this measurement the speed of execution is calculated. The execution time of doing
only the handshaking between the host and the \textit{RC-1000} system (i.e. with no computations)
is approximately $132 \mu Sec$. The following code section shows the part probing for the
execution time in \textit{C++}.

{\singlespace  {\small
\begin{verbatim}
    LONGLONG llFrequency, llTimeBefore, llTimeAfter, llTimeTaken;

    double dTimeTakenInSeconds;
    double TimeResults[TestingLoop];
    .
    .
    .
    if (!QueryPerformanceFrequency ((LARGE_INTEGER *) &llFrequency )){
        return 0;}

    //////////////////////////////////////////////////////////
    QueryPerformanceCounter ((LARGE_INTEGER *)&llTimeBefore);

    PP1000WriteControl(Handle, 123);
    PP1000ReadStatus(Handle, &ReturnVal);

    QueryPerformanceCounter ((LARGE_INTEGER *)&llTimeAfter);
    //////////////////////////////////////////////////////////

    llTimeTaken = llTimeAfter - llTimeBefore;
    dTimeTakenInSeconds = ((double)llTimeTaken) / ((double)llFrequency);
    TimeResults = dTimeTakenInSeconds;
\end{verbatim}}}

\end{comment}
%%%%%%%%%%%%%%%%%%%%%%%%%%%%%%%%%%%%%%%%%%%%%%%%%%%%%%%%%%%%%%%%%%%%%%%%%%%%%%%%%%%

The information about the hardware area occupied by a design, i.e. number of Slices used after
placing and routing the compiled code, is determined by the \textit{ISE} place and route tool.
Using the same tool we are able to get more detailed statistics about our compilation. In the
current investigation the only used metrics are the number of Slices and the Total Equivalent Gate
Count for a design.

%The used metrics are summarized in Table~\ref{Metrics2}.

%%%%%%%%%%%%%%%%%%%%%%%%%%%%%%%%%%%%%%%%%%%%%%%%%%%%%%%%%%%%%%%%%%%%%%%%%%%%%%%%%%%%%%%%%%%%%%%%%

\section{A New Approach for developing a Matrix Multiplication Algorithm} \label{App}
Many parallel implementations of matrix multiplication have been
investigated in the literature. Although this algorithm is simple
and it has a long history, the continuous advancement in computer
architectures made the study of this algorithm very interesting.
Many matrix multiplication algorithms were suggested for parallel
implementation. Horowitz and Zorat in \cite{MM6} and Hake in
\cite{MM7} suggested a recursive divide-and-conquer solution. Fox
et al. in \cite{MM8} and Canon in \cite{MM5} presented different
ways this algorithm could be developed for a mesh topology. Other
implementations were discussed in \cite{MM1,MM2,MM3,MM4}.

An important requirement for any parallel hardware computation is
the proper use of available computing resources. This is done
either to minimise overall computation time, to minimise the chip
area, or to compromise between these two goals. With the
development model in hand, design flexibility is one of the main
advantages granted. Accordingly, five refined designs from the
functional specification of the standard matrix multiplication
algorithm are presented. These designs vary between high-speed
implementation with expensive use of resources, and lower speed
implementation with less use of resources. The development of the
matrix multiplication algorithm is presented in the following
sections.

The development will start by formalizing the matrix
multiplication algorithm. The functional specification will favor
the use of the predefined high-order functions. This functional
specification will be the source for different refinements with
different degrees of parallelism described using \textit{CSP}
notation. The created parallel designs will be used in the section
Section~\ref{RHI} as the basis of the code written in
\textit{Handel-C}.

%%%%%%%%%%%%%%%%%%%%%%%%%%%%%%%%%%%%%%%%%%%%%%%%%%%%%%%%%%%%%%%%%%%%%%%%%%%%%%%%%

\section{Formal Functional Specification} \label{FFS}

An informal definition of the problem considers that the multiplication of two matrices
\textit{ass} and \textit{bss} produces the matrix \textit{css} whose elements, $c_{ij}(0 \leq i <
n, 0 \leq j < k)$ are computed as follows:

$c_{ij} = \sum_{t = 0}^{m - 1}a_{i,t}b_{t,j}$

\noindent where \textit{ass} is an $n \times m$ matrix and \textit{bss} is an $m \times k$ matrix.
Items \textit{a} and \textit{b} correspond to elements from matrices \textit{ass} and
\textit{bss}, respectively.

Generally, partitioning can be done very easily with matrix multiplication, where each matrix is
divided into sub-matrices that can be manipulated as if they were a single matrix element
\cite{MM8}. This method is used to divide matrices with large dimensions to suit the expected
limited capability of the available computer.

Turning our attention to the formalisation of the algorithm. A functional specification of matrix
multiplication is formulated as a function \textit{mmult} that takes two matrices as inputs and
returns a matrix as a result. In this definition, we assume the first matrix is represented as a
list of rows and the second matrix is represented as a list of columns.

{\singlespace  {\small \begin{verbatim}
 mmult :: [[Int]] -> [[Int]] -> [[Int]]
 mmult ass bss = map (vmmult ass) bss

 vmmult :: [[Int]] -> [Int] -> [Int]
 vmmult ass bs = map (scalarp bs) ass

 scalarp :: [Int] -> [Int]-> Int
 scalarp  as  bs = sum (zipwithmul as bs)

 sum :: [Int] -> Int
 sum  rs = foldrl (+)  rs

 zipwithmul :: [Int] -> [Int] -> [Int]
 zipwithmul  as  bs = zipwith (*) as bs
\end{verbatim}}}

The suggested algorithm for multiplying two matrices is done by mapping (using the higher-order
function \textit{map}) the function (\textit{vmmult ass}) to all vectors in \textit{bss}. This
function is the multiplication of a vector with a matrix. It takes two inputs a matrix (list of
lists) \textit{ass} and a vector (list) \textit{bs} and returns a list \textit{cs} (a column in
the resulting matrix).

In turn, \textit{vmmult} maps the function (\textit{scalarp bs}) over the list of lists
\textit{ass}. The function \textit{scalarp} defines the scalar product of two vectors. The inputs
to this function are two lists \textit{as} and \textit{bs}. The higher-order function
\textit{zipWith} is used in the function \textit{zipwithmul} to zip the inputs with
multiplication, then the function \textit{sum} is employed to fold the already zipped lists with
addition. The output of this composition is an element from the resultant matrix.

According to this specification, the implementation under \textit{HUGs98} \textit{Haskell}
compiler is tested at the unit, component and integration levels.

%%%%%%%%%%%%%%%%%%%%%%%%%%%%%%%%%%%%%%%%%%%%%%%%%%%%%%%%%%%%%%%%%%%%%%%%%%%%%%%%%

\section{Algorithm Refinements} \label{AF}

Clearly, parallelism is not a part of the starting specification of the stated problem. Typically,
parallelism and communications are introduced at this stage in the development for the sole
purpose of capturing functionally equivalent, but parallel, designs.

Applying the provably correct refinement rules, the previous specification is refined to
\textit{CSP} as a middle stage towards hardware realisation. The capability of doing different
data refinements implies the availability of various designs. Whereby, each design would have
different characteristics and levels of parallelism.

In the following refinements, five designs are presented. The first design is a data-parallel
design, while, the second design is a stream-based design. The third and fourth designs addresses
refinement to pipelined parallelism using function decomposition strategy. The last design is a 2D
pipelined design with an extension leading to a systolic implementation of the problem.

For more clarification of the used terms we recall the following informal definitions. A
data-parallel design replicates the same processes in order to compute for different data inputs.
Commonly, a Single Program Multiple Data (SPMD) approach is used in data-parallel models, where
data are distributed across processors. In pipelined computations, a program is divided into a
series of tasks that have to be completed consecutively. Accordingly, these tasks are executed by
separate pipeline stages. The pipe stages stream data from stage to stage to form the required
computation. A stream-based design eliminates some replication (in data-parallel designs) or some
stages (in pipelined designs) and it processes the input and output as streams of data. Systolic
arrays are another parallel computing architecture. It is best described by analogy with the
regular pumping of blood by the heart. In systolic arrays, processors are arranged in an array
where data flow across the array elements between neighbours. For instance, a process firstly
takes in data from one or more neighbours (North and West). Secondly, the process manipulates the
input data. Finally, the process outputs results in the opposite direction (South and East).

\subsection{First Design - Data Parallelism}

Recalling the high-level specification of \textit{mmult}:

{\singlespace  {\small
\begin{verbatim}
 mmult :: [[Int]] -> [[Int]] -> [[Int]]
 mmult ass bss = map(vmmult ass) bss
\end{verbatim}}}

In this design we consider the refinement of the input \textit{bss} and the output \textit{css} as
vectors of items of size k, where each item is a list.

$mmult(ass) :: \lfloor[Int]\rfloor_{k} \rightarrow \lfloor[Int]\rfloor_{k}$

The \textit{CSP} implementation of the functional specification of the matrix multiplication
algorithm \textit{mmult} realises this function as a process \textit{MMULT}. The \textit{CSP}
process \textit{MMULT} is the parallel execution of k-copies of \textit{VMMULT} (the refinement of
\textit{vmmult}). This description is implemented using \textit{VMAP} the vector setting of the
higher-order function \textit{map}. Therefore, it is the interleaving with renaming of the process
\textit{VMMULT}  for all columns of \textit{bss}:

$mmult(ass) \sqsubseteq MMULT(ass)$
$MMULT(ass) = VMAP_{k}(VMMULT(ass))$

\begin{figure} [htpb]
	\begin{center}
		\includegraphics [scale = 1]%[height=2in,width=2in,angle=0]
		{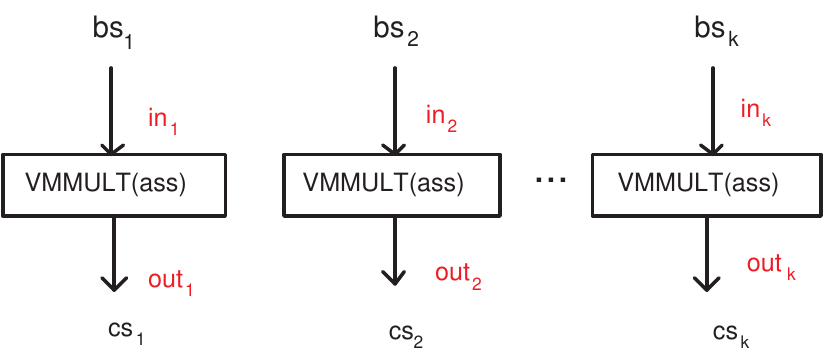}
		\caption{The process MMULT}
		\label{fig1}
	\end{center}
\end{figure}

The list \textit{ass} is passed as an argument to each of the processes
\textit{VMMUL}(\textit{ass}) in the above design. This design can be pictured as in
Figure~\ref{fig1}.

The list \textit{ass} could be explicitly passed to the process \textit{VMMULT} by exploiting the
following algebraic identity:

$VMMULT(ass)= PRD(ass) \rhd VMMULT$

\begin{figure} [htpb]%[f]
	\begin{center}
		\includegraphics [scale = 1]%[height=2in,width=2in,angle=0]
		{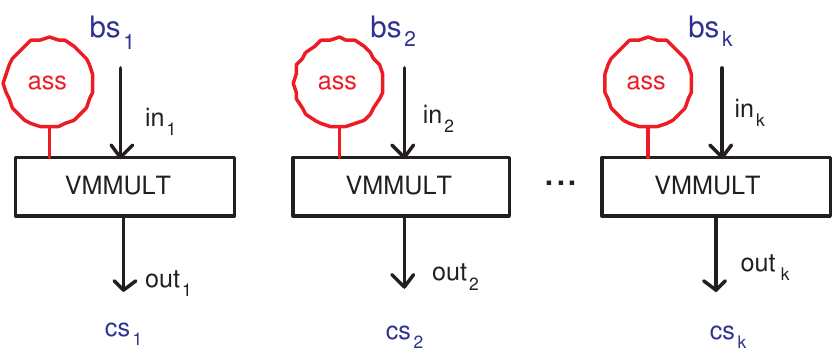}
		\caption{The process MMULT, an alternative implementation}
		\label{fig2}
	\end{center}
\end{figure}

\begin{figure} [htpb]%[f]
	\begin{center}
		\includegraphics [scale = 1]%[height=2in,width=2in,angle=0]
		{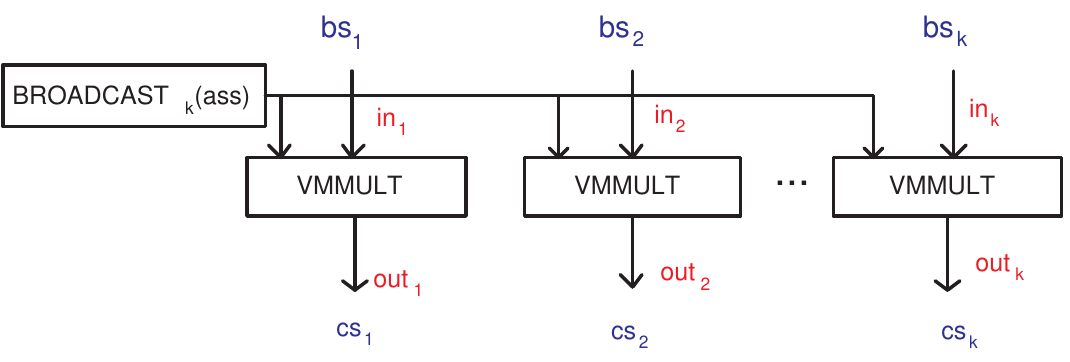}
		\caption{The process MMULT, optimised implementation}
		\label{fig3}
	\end{center}
\end{figure}

The effect of applying this step to the previous design is visualised in Figure~\ref{fig2}. In
this version, the list \textit{ass} is locally produced and fed to each process \textit{VMMULT} in
the vector. The effect of having \textit{k} parallel copies of \textit{PRD}(\textit{ass})
communicating with \textit{k} instances of \textit{VMMULT} can be achieved by factorising the
process \textit{PRD}(\textit{ass}) and broadcasting its output to the relevant processing elements
in the network. Applying this rule will result in a semantically equivalent version of
\textit{MMULT} which has a different layout, this is shown in Figure~\ref{fig3}.

Now we turn our attention to the refinement of the function \textit{vmmult}.

{\singlespace  {\small
\begin{verbatim}
 vmmult::[Int] -> [[Int]] -> [Int]
 vmmult bs ass = map (scalarp bs) ass
\end{verbatim}}}

Clearly, \textit{vmmult (bs)} is a map pattern. Since \textit{map} has two different
implementations, we will consider them in turn in this and the next designs. In this design, the
\textit{CSP} implementation realises the function \textit{vmmult} as a process \textit{VMMULT}
with a vector of items \textit{ass}=$\lfloor\textit{as}_{1}, \textit{as}_{2}, ...
\textit{as}_{n}\rfloor$ as input and a vector of items \textit{cs}=$\lfloor\textit{c}_{1},
\textit{c}_{2}, ... \textit{c}_{n}\rfloor $ as output. By refining \textit{scalarp} into
\textit{VSCALARP}, the \textit{CSP} implementation of \textit{vmmult (bs)} is again the off the
shelf refinement of a vector map:

$vmmult(bs) :: \lfloor [Int] \rfloor_{m} \rightarrow [Int]_{n}$

$VMMULT(bs) = VMAP_{n}(VSCALARP(bs))$

By appealing to the same technique already used in the refinement of \textit{VMMULT} we get:

$VSCALARP(bs) = PRD(bs) \rhd VSCALARP$

This leads to a new design of\textit{VMMULT(bs)}:

$VMAP_{n}(PRD(bs) \rhd VSCALARP)$

$BROADCAST_{n}(bs)\rhd VMAP_{n}(VSCALARP(bs))$

\begin{figure} [htpb]%[f]
	\begin{center}
		\includegraphics [scale = 1] %[height=2in,width=4in,angle=0]
		{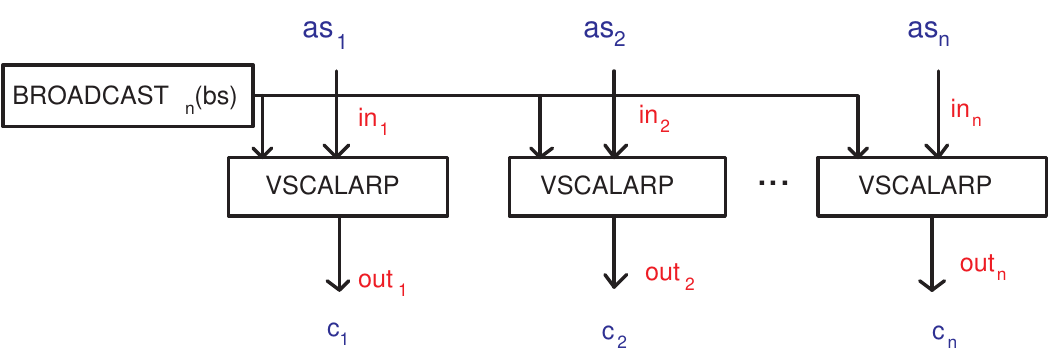}
		\caption{The process VMMULT}
		\label{fig4}
	\end{center}
\end{figure}

Figure~\ref{fig4} shows the process \textit{VMMULT}. This step also shows clearly the replication
of the process \textit{VSCALARP}, which is an indicator for a later replication in the hardware
implementation.

\begin{figure} [htpb]
	\begin{center}
		\includegraphics [scale = 0.8]
		{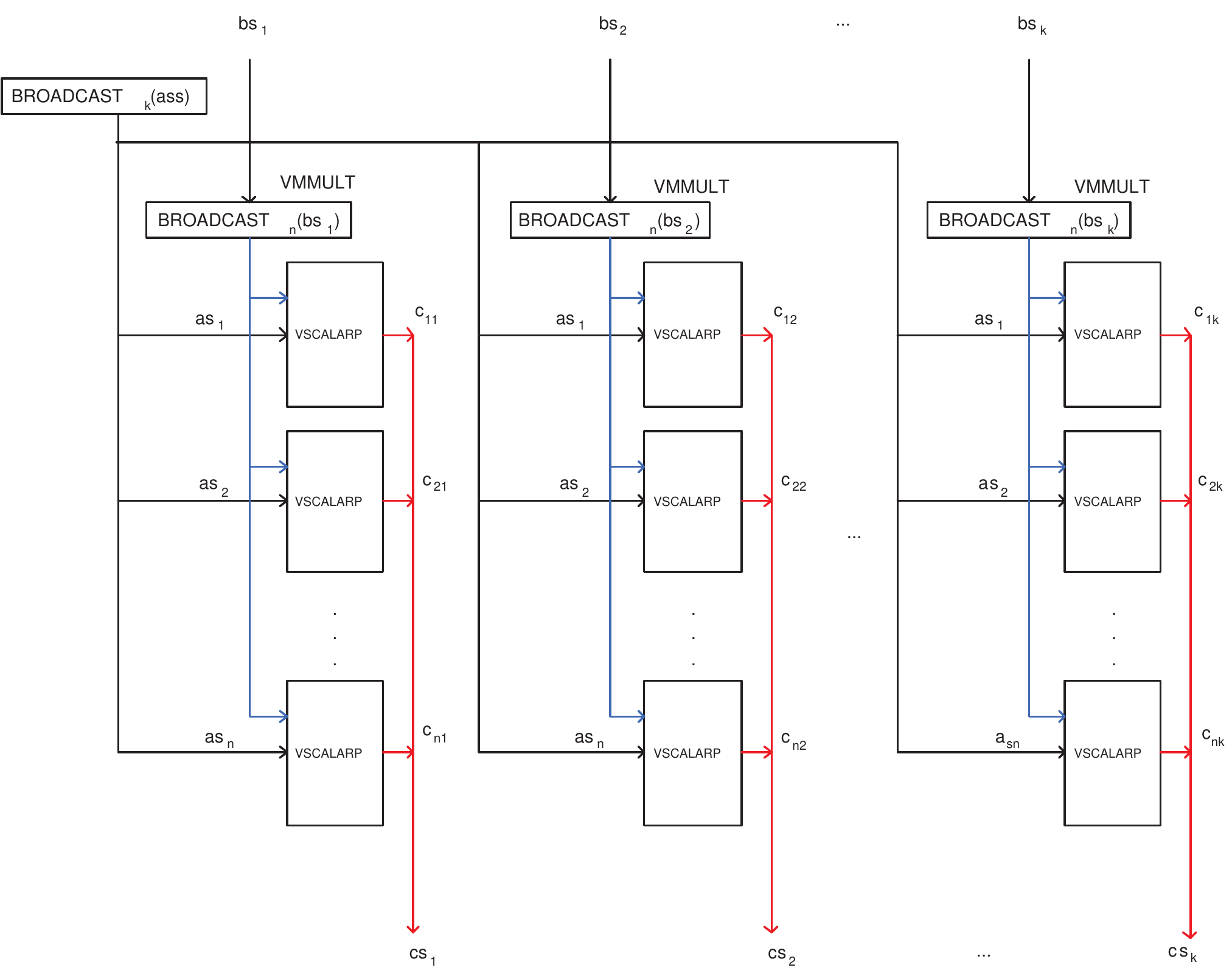}
		\caption{The process MMULT, data-parallel design}
		\label{fig5}
	\end{center}
\end{figure}

Figure~\ref{fig5} expands the main building block in Figure~\ref{fig3} by corresponding
configuration in Figure~\ref{fig4}. This gives a two dimensional visualisation of the process
\textit{MMULT} as a data parallel implementation.

The next step is to present the building block \textit{VSCALARP} corresponding to a \textit{CSP}
refinement of the function:

{\singlespace  {\small
\begin{verbatim}
 scalarp :: [Int] -> [Int] -> Int
 scalarp  as  bs = sum (zipwithmul as bs)
\end{verbatim}}}

This function can be refined as the piping of two processes \textit{VZIPWITH} and \textit{VFOLD}
corresponding to refinements of  the functions \textit{zipwithmul} and \textit{sum} respectively.

$scalarp:: \lfloor Int \rfloor_{m} \rightarrow \lfloor Int \rfloor_{m}\rightarrow Int$

$VSCALARP = VZIPWITH_{m}(MUL)>>_{m}VFOLD_{m}(ADD)$

\begin{figure} [htpb]
	\begin{center}
		\includegraphics [scale = 1]%[height=2in,width=2in,angle=0]
		{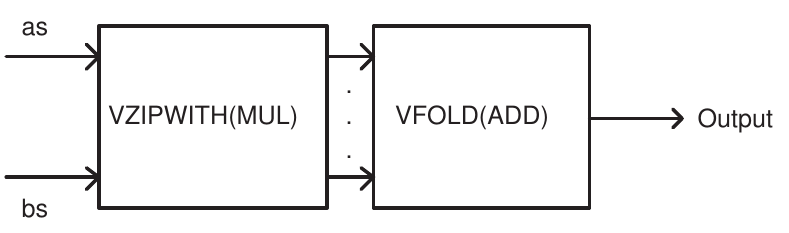}
		\caption{VSCALARP as a piping of two processes}
		\label{fig6}
	\end{center}
\end{figure}

This description is depicted in Figure~\ref{fig6}.

For completeness, the \textit{CSP} implementations of the simple addition and multiplication
functions are:

$ADD=(in_{1}? a \rightarrow SKIP \interleave in_{2}? b \rightarrow SKIP); out!(a+b)$

$MUL = (in_{1}? a \rightarrow SKIP \interleave in_{2}? b\rightarrow SKIP)$;$out!(a \times b)$

%%%%%%%%%%%%%%%%%%%%%%%%%%%%%%%%%%%%%%%%%%%%%%%%%%%%%%%%%%%%%%%%%%%%%%%%%%%%%%%%%

\subsection{Second Design - Streaming I/O} For this design the
refinement of \textit{mmult} is not changed. The change will
appear in the refinement of the function \textit{vmmult}. Recall
the formal specification of this function:

{\singlespace  {\small
\begin{verbatim}
 vmmult::[Int] -> [[Int]] -> [Int]
 vmmult bs ass = map (scalarp bs) ass
\end{verbatim}}}

At this point, the input list \textit{ass} is viewed as a stream of values \textit{ass}=
$\langle\textit{as}_{1}, \textit{as}_{2}, ... \textit{as}_{n}\rangle $ and the output list as a
stream of values as well. The \textit{CSP} refinement of\textit{ vmmult(bs)} is directly obtained
from the off-the-shelf stream-based implementation of the higher-order function map:

$vmmult(bs) :: \langle [Int] \rangle \rightarrow \langle Int \rangle$

$VMMULT(bs) = MAP (VSCALARP(bs))$

\begin{figure} [htpb]
	\begin{center}
		\includegraphics [scale = 0.85]%[height=2in,width=2in,angle=0]
		{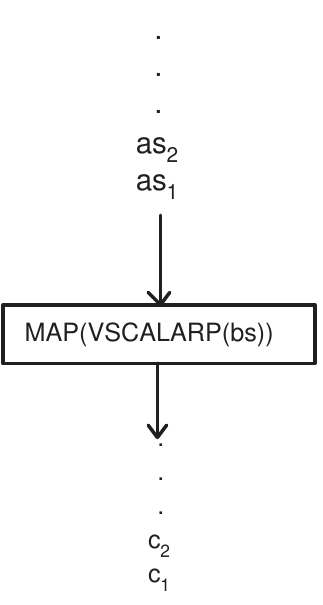}
		\caption{The process VMMULT the input and output refined as streams of values}
		\label{fig7}
	\end{center}
\end{figure}

Figure~\ref{fig7} shows the new version of the process \textit{VMMULT}. This step also shows
clearly that there is no more replication of the process\textit{ VSCALARP}, which is an indicator for the later reduction in use of hardware resources.

\begin{figure} [htpb]
	\begin{center}
		\includegraphics [scale = 0.85]
		{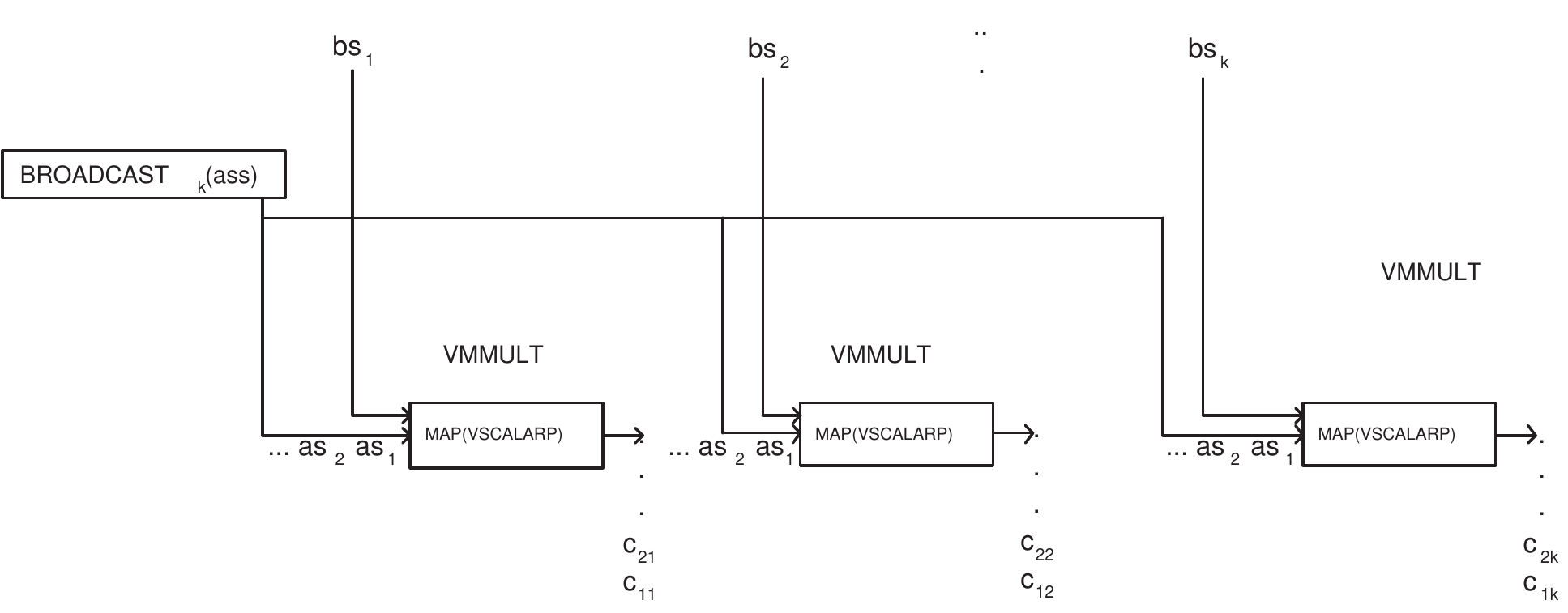}
		\caption{The process MMULT the input and output refined as streams of values}
		\label{fig8}
	\end{center}
\end{figure}

Keeping the refinement of the remaining functions the same, \textit{MMULT} process looks as in
Figure~\ref{fig8}.

%%%%%%%%%%%%%%%%%%%%%%%%%%%%%%%%%%%%%%%%%%%%%%%%%%%%%%%%%%%%%%%%%%%%%%%%%%%%%%%%%

\subsection{Third Design - Pipelining} Demonstrating the
refinement to pipelined parallelism is the purpose of this design.
Generally, this kind of parallelism is a very effective means for
achieving efficiency in numerous algorithms. Usually, pipelined
parallelism is much harder to detect than data parallelism.
Accordingly, the function decomposition strategy, found in
\cite{8} and recalled in Section~\ref{Decmpstn} is used. This
strategy aims at exhibiting pipelined parallelism in functional
programs. According to the decomposition rule the definition of
the function \textit{mmult} is pipelined.

{\singlespace  {\small
\begin{verbatim}
 mmult  ass  bss  = map (vmmult ass) bss

 vmmult [] bs = []
 vmmult  (as : ass) bs = vscalarp as  bs : vmmult ass  bs
\end{verbatim}}}

The recursive function in this case is \textit{vmmult}. The value to be passed to the next stage
of the pipe is a tuple. Its first is the input vector and its second is result of applying
\textit{vscalarp} on the input vector from matrix \textit{bss} and the present argument from
matrix \textit{ass}. According to the decomposition rule, the efficient implementation of
\textit{mmult} as a pipelined network of \textit{CSP} processes can be as follows:

$MMULT = PRD(bss) \rhd ((\gg)/ (MAP \circ f')\ast [as_{n},as_{n-1},...,as_{0}])$

$MAP(initial)= \mu Z$ \ $\bullet$ \ \ \ $left?eot \rightarrow right!eot \rightarrow SKIP$
\\
$|$
\\
$left?x \rightarrow right! \langle bs, [ ] \rangle \rightarrow Z $

$MAP(f'$ \ $as) = \mu Z$ \ $\bullet$ \ \ $left?eot \rightarrow right!eot \rightarrow SKIP$
\\
$|$
\\
$left?\langle bs, y \rangle \rightarrow right! \langle bs, y \pp (vscalarp$ \ $as$ \ $bs) \rangle
\rightarrow Z $

$MAP(final)= \mu Z$ \ $\bullet$ \ \ \ $left?eot \rightarrow right!eot \rightarrow SKIP$
\\
$|$
\\
$left? \langle bs, y \rangle \rightarrow right!y \rightarrow Z$
\\

\begin{figure} [htpb]
	\begin{center}
		\includegraphics [scale = 0.7]
		{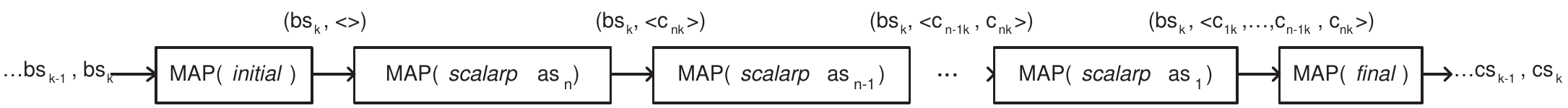}
		\caption{The process MMULT as a pipelined network, third design}
		\label{fig9}
	\end{center}
\end{figure}

The decomposed pipelined network is shown in Figure~\ref{fig9}. In this design, the matrix
\textit{bss} is input to the network as a stream of vectors (columns)  $\langle\textit{bs}_{1},
\textit{bs}_{2}, ... \textit{bs}_{k}\rangle $. The matrix \textit{ass} vectors (row by row) are
produced in the pipe stages. The result is considered as a stream of streams
$\langle\textit{cs}_{1}, \textit{cs}_{2}, ... \textit{cs}_{k}\rangle $ The first result to appear
from the network is the output stream  (column) $\langle \textit{cs}_{k}\rangle $ corresponding to
the first input vector \textit{bs$_{k}$}. This design is independent from the size of \textit{k} a
dimension of \textit{bss} and \textit{css}.

%%%%%%%%%%%%%%%%%%%%%%%%%%%%%%%%%%%%%%%%%%%%%%%%%%%%%%%%%%%%%%%%%%%%%%%%%%%%%%%%%

\subsection{Fourth Design - Pipelined Turnout Stages} 
\begin{figure} [htpb]
	\begin{center}
		\includegraphics [scale = 1]
		{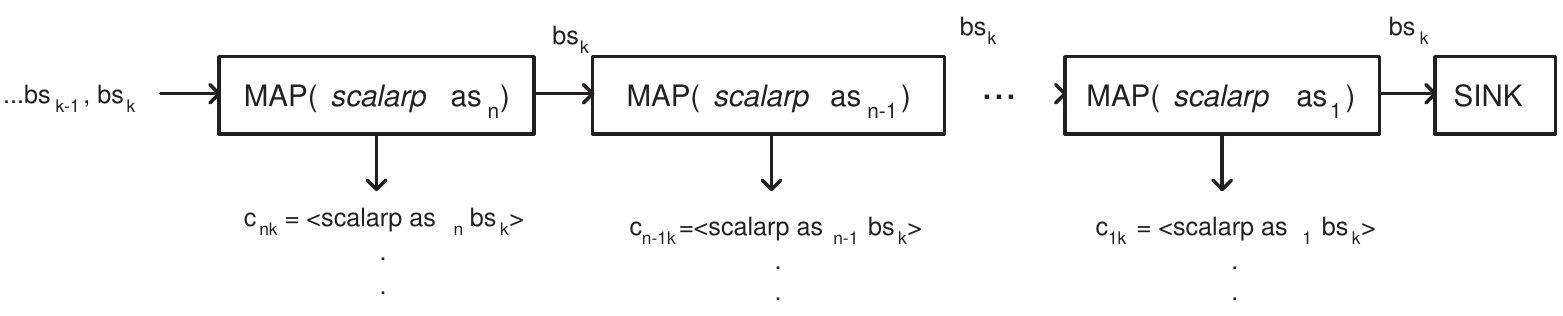}
		\caption{The process MMULT as a pipelined network, fourth design}
		\label{fig10}
	\end{center}
\end{figure}

This design
makes use of an optimisation of the previous design. In this case,
the input matrix\textit{ bss} is refined as a stream of vectors
$\langle\textit{bs}_{1}, \textit{bs}_{2}, ...
\textit{bs}_{k}\rangle $ and the matrix \textit{ass} is refined as
arguments in the pipeline stages. The output from each stage is
turned out as a result, besides forwarding the input from
\textit{bss} to the next stage. Thus, the output from this
pipeline is a vector of streams as shown in Figure~\ref{fig10}.
Note that, this design also doesn't depend on the size of k - the
dimension in \textit{bss} and \textit{css}. The \textit{CSP}
implementation is as follows:

$MMULT = PRD(Bss)((\gg)/ (MAP \circ f')\ast [as_{n},as_{n-1},...,as_{0}]) \parallel SINK $

$MAP(f'$ \ $as) = \mu Z$ \ $\bullet$ \ \ \ $left?eot \rightarrow right!eot \rightarrow SKIP$

\ \ \ \ \ \ \ \ \ \ \ \ \ \ \ \ \ \ \ \ \ \ \ \ \ \ \ \ \ \ \ $|$

\ \ \ \ \ \ \ \ \ \ \ \ \ \ \ \ \ \ \ \ \ \ \ \ \ \ \ \ \ \ \ \ \ $left?bs \rightarrow
down!(vscalarp$ \ $as$ \ $bs) \rightarrow right!bs \rightarrow Z $

%%%%%%%%%%%%%%%%%%%%%%%%%%%%%%%%%%%%%%%%%%%%%%%%%%%%%%%%%%%%%%%%%%%%%%%%%%%%%%%%%

\subsection{Multilevel Pipelined Design} 
\begin{figure} [htpb]
	\begin{center}
		\includegraphics [scale = 0.9]
		{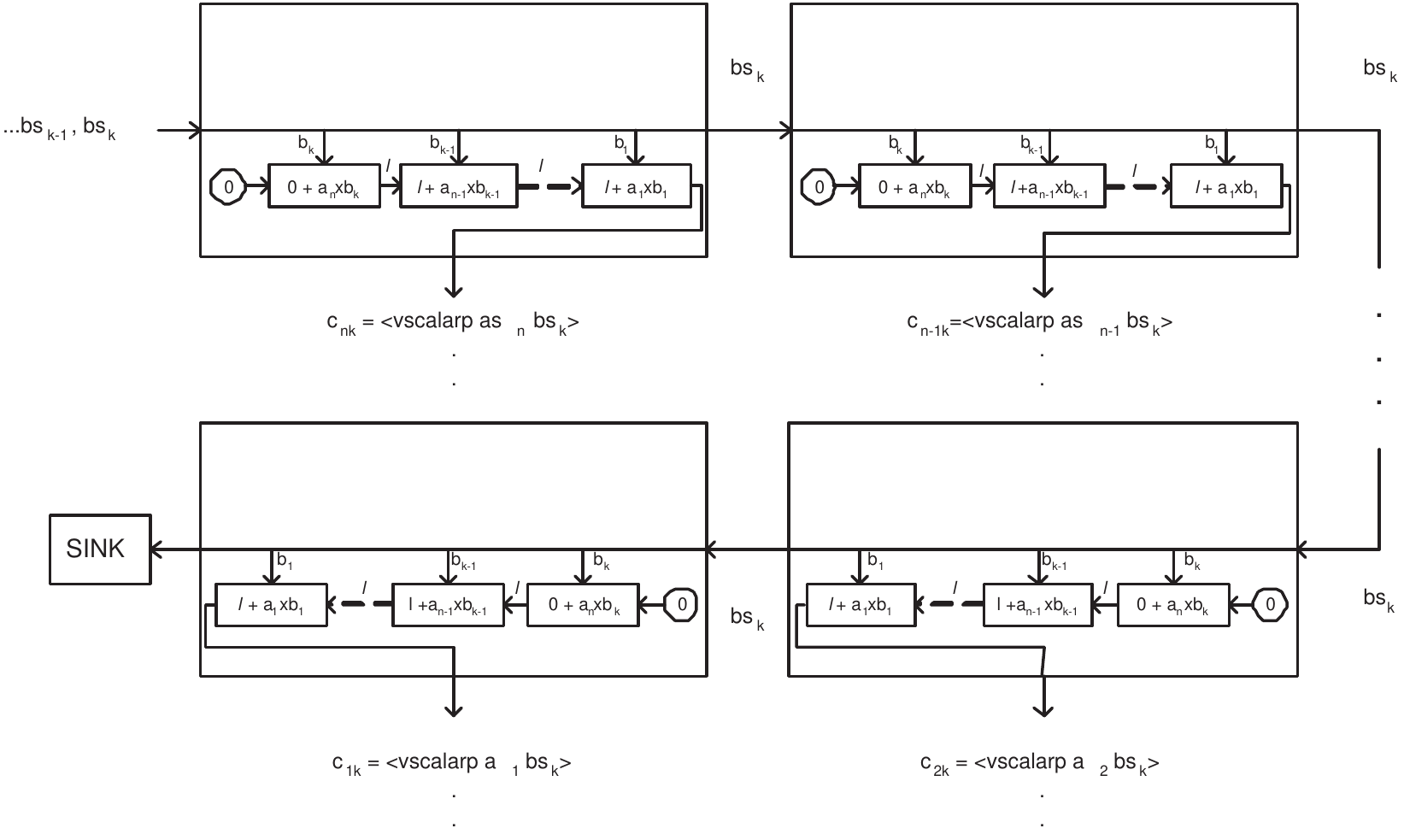}
		\caption{The process MMULT as 2D network design}
		\label{MMULT72D}
	\end{center}
\end{figure}

This design applies the
function decomposition strategy for pipelined parallelism on two
levels. The first level is decomposing the vector matrix
multiplication into a pipeline of processes performing the scalar
product of two vectors, this is similar to the fourth design. An
addition is made by pipelining the scalar product routine creating
a second level pipeline. The final structure of the suggested
design is multilevel pipelines realising the matrix multiplication
algorithm. The decomposition of the process \textit{VSCALARP} is
done in a similar manner (See Figure~\ref{MMULT72D}):
\\

$VSCALARP = PRD(0) \rhd ((\gg)/ (MAP \circ f')\ast [a_{m},a_{m-1},...,a_{0}])$

$MAP(f'$ \ $a) = \mu Z$ \ $\bullet$ \ \ \ $left?eot \rightarrow right!eot \rightarrow SKIP$

\ \ \ \ \ \ \ \ \ \ \ \ \ \ \ \ \ \ \ \ \ \ \ \ \ \ \ \ \ \ $|$

\ \ \ \ \ \ \ \ \ \ \ \ \ \ \ \ \ \ \ \ \ \ \ \ \ \ \ \ \ \ $left? x \rightarrow up? b \rightarrow
right! (x + (a \times b)) \rightarrow Z $

Thus \textit{MMULT} implementation is as follows:

$MMULT = PRD(Bss)((\gg)/ (MAP \circ f')\ast [as_{n},as_{n-1},...,as_{0}]) \parallel SINK $

$MAP(f'$ \ $as) = \mu Z$ \ $\bullet$ \ \ \ $left?eot \rightarrow right!eot \rightarrow SKIP$

\ \ \ \ \ \ \ \ \ \ \ \ \ \ \ \ \ \ \ \ \ \ \ \ \ \ \ \ \ \ \ $|$

\ \ \ \ \ \ \ \ \ \ \ \ \ \ \ \ \ \ \ \ \ \ \ \ \ \ \ \ \ \ \ \ \ $left?bs \rightarrow (PRD(bs)
\rhd VSCALARP(as)[down/right]);$

\ \ \ \ \ \ \ \ \ \ \ \ \ \ \ \ \ \ \ \ \ \ \ \ \ \ $right ? bs \rightarrow Z $

\newpage

\begin{figure} [htpb]
	\begin{center}
		\includegraphics [scale = 1]
		{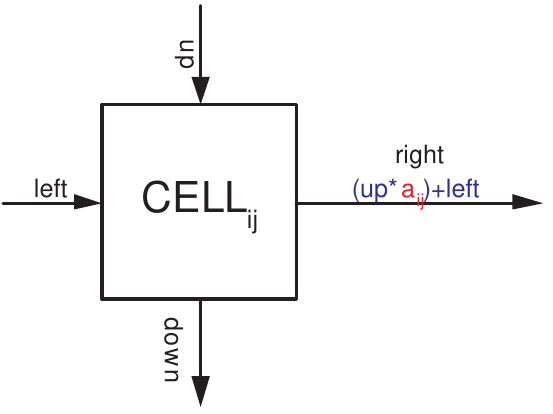}
		\caption{Basic cell}
		\label{fig11}
	\end{center}
\end{figure}

An optimisation of this design would lead to a systolic solution. The main idea of the change is
to enable the communication between parallel \textit{VSCALARP} stages in \textit{MMULT}. A
\textit{VSCALARP} is to be the parallel execution of basic cells, each responsible of forwarding
down the upper input from \textit{bss}. The cell is also responsible for doing its part for the
scalar product computation. This part is done by outputting to the right the result of adding the
left input to the multiplication of the upper input \textit{b} (from \textit{bss}) with the item
\textit{a} (from \textit{ass}). The basic process is called \textit{CELL} (See Figure~\ref{fig11})
and defined as:

$CELL (a)= PRD(a) \rhd (up ? u \rightarrow left? l  \rightarrow right!(u*a + l) \rightarrow down!u
)$

Then, \textit{VSCALARP} is implemented as:

$VSCALARP(as) = PRD(0)\rhd$

$(\parallel_{i=1}^{i=m}CELL(as[i])[d_{i}/left, d_{(i+1)}/right, e_{i}/up, e_{(i+1)}/down])$

Thus, \textit{MMULT} implementation is as follows:

$MMULT = PRD(Bss)((\gg)/ (MAP \circ f')\ast [as_{n},as_{n-1},...,as_{0}]) \parallel SINK$

$MAP(f'$ \ $as) = \mu Z$ \ $\bullet$ \ \ \ $up?eot \rightarrow down!eot \rightarrow SKIP$

\ \ \ \ \ \ \ \ \ \ \ \ \ \ \ \ \ \ \ \ \ \ \ \ \ \ \ \ \ \ \ $|$

\ \ \ \ \ \ \ \ \ \ \ \ \ \ \ \ \ \ \ \ \ \ \ \ \ \ \ \ \ \ \ \ \ $VSCALARP(as)[right/d_{m}]; Z $

\begin{figure} [htpb]
	\begin{center}
		\includegraphics [scale = 0.9]
		{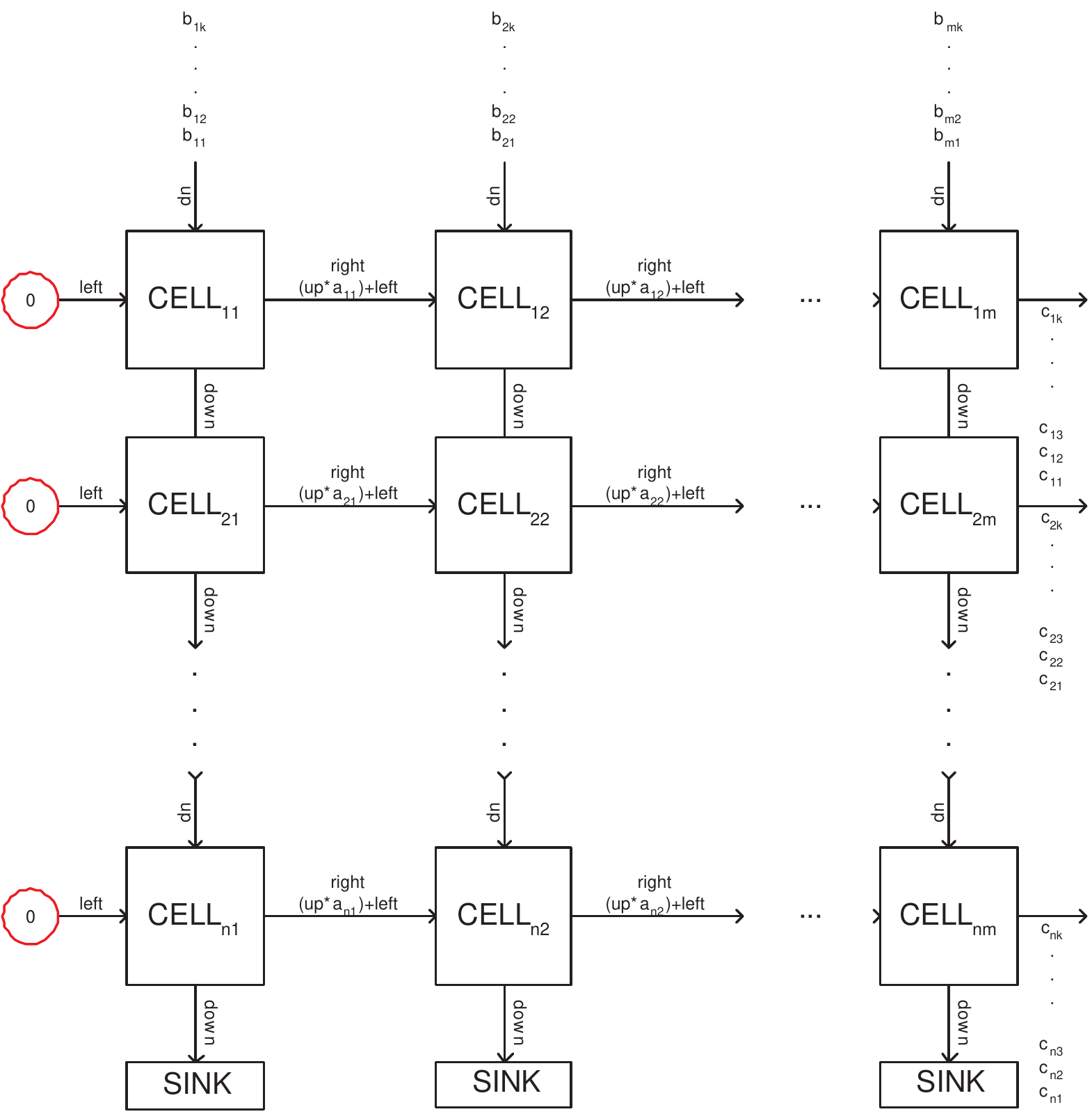}
		\caption{The process MMULT as a systolic network}
		\label{fig12}
	\end{center}
\end{figure}

This implementation is depicted in Figure~\ref{fig12}.

For the sake of giving a similar design, an intuitive \textit{CSP} implementation of the matrix
multiplication algorithm shown in Figure~\ref{fig12} is as follows:

$CELL_{(i,j)} (ass[i,j])= \mu X  \bullet PRD(ass[i,j]) \rhd ((up ? u \rightarrow down!u
\rightarrow (SKIP \lessdot u = eot \gtrdot left? l \rightarrow right!(u*ass[i,j] + l)) \rightarrow
SKIP \rightarrow X)$

The matrix multiplication process \textit{MMULT} is then implemented as:

$MMULT = BROADCAST_{n}(0)[d/out] \rhd$

$(\parallel_{i=1}^{i=n} (\parallel_{j=1}^{j=m}CELL(i, j)[b_{ij}/lwft, b_{i(j+1)}/right, e_{ij}/up
, e_{(i+1)j}/down]))$

Finally, these designs depend only on the dimensions \textit{n} and \textit{m} from \textit{ass}.

%%%%%%%%%%%%%%%%%%%%%%%%%%%%%%%%%%%%%%%%%%%%%%%%%%%%%%%%%%%%%%%%%%%%%%%%%%%%%%%%%

\section{Reconfigurable Hardware Implementation}  \label{RHI}

As a stage in the development model, \textit{Handel-C} code
follows the refined \textit{CSP} implementation of the presented
designs. The targeted circuit implementation is to have the same
topology of the communicating processes shown in the refinement
section. In the following subsections, some pieces of code taken
from the five different designs are presented.

%%%%%%%%%%%%%%%%%%%%%%%%%%%%%%%%%%%%%%%%%%%%%%%%%%%%%%%%%%%%%%%%%%%%%%%%%%%%%%%%%

\subsection{First Design - Data Parallelism}

From first design, we recall the \textit{CSP} implementation of
the process \textit{MMULT}:

$MMULT(ass) = VMAP_{K}(VMMULT(ass))$

The code corresponding to the process \textit{MMULT} is the macro \textit{MatrixMult} with
\textit{bss} as an input and \textit{css} as an output. This macro is vector mapping of the macro
\textit{VectMatrixMult}, where the vector of vectors \textit{ass} is internally produced. Recall
that the macro \textit{VMap} is the implementation of the higher-order process \textit{VMAP}. In
this case, \textit{VMap} works by distributing the matrix \textit{bss} for each process
\textit{VectMatrixMult}, where a vector from \textit{css} will be the output.

{\singlespace  {\small
\begin{verbatim}
 macro proc MatrixMult (bss, css, n){
    VMap(bss, css, n, VectMatrixMult);}
\end{verbatim}}}

Then, the macro \textit{VectMatrixMult} implements the process \textit{VMMULT}, refined as:

$VMMULT(ass)= PRD(ass) \rhd VMMULT$

This applies again \textit{VMap} calling the macro \textit{VScalarP} (the implementation of the
process VSCALARP) for each vector in \textit{ass}. At his point, \textit{bss} is sinked as it will
be later internally produced in \textit{VScalarP}.

{\singlespace  {\small
\begin{verbatim}
 macro proc VectMatrixMult (bss, cs) {
    VectorOfVectorsOfItems(ass, n, m, Int);
    SinkVectorOfVectorsOfItems(bss, n, m, sink);
    par {
    ProduceVectorOfVectorsOfItems(ass, n, m, tempAss);
    VMap(ass, cs, n, VScalarP);}}
\end{verbatim}}}

The macro \textit{VScalarP} implements the \textit{CSP} process \textit{VSCALARP}:

$VSCALARP(bs) = PRD(bs) \rhd VSCALARP$

The internal production for a  vector from \textit{bss} is done according to an index. This allows
producing a different vector \textit{bs} from \textit{bss} for each execution of
\textit{VScalarP}.

{\singlespace  {\small
\begin{verbatim}
 macro proc VScalarP(as, outputItem, index){
    VectorOfItems(internalV, m, Int);
    VectorOfItems(bs, m, Int);
    par {
    ProduceVectorOfItems(bs, m, tempBss[index]);
    VZipWith(m, as, bs, internalV, Multiplication);
    VFoldR (internalV, outputItem, m, Addition, 0);}}
\end{verbatim}}}

The macros Addition and Multiplication corresponding to the processes \textit{ADD} and
\textit{MUL} are implemented as:

{\singlespace  {\small
\begin{verbatim}
 macro proc Multiplication(xItem, yItem, output) {
    Int x,y;
    xItem.Channel ? x;
    yItem.Channel ? y;
    output.Channel ! (x * y);}

 macro proc Addition(xItem, yItem, output) {
    Int x,y;
    xItem.Channel ? x;
    yItem.Channel ? y;
    output.Channel ! (x + y);}
\end{verbatim}}}

Running the above implementation is done practically by producing \textit{bss} and storing
\textit{css} from/to a buffer. The \textit{RC-1000} board internal \textit{SRAMS} are used as the
input and output buffers. The main macro call that runs the code implementing the first design for
the matrix multiplication algorithm is as follows:

{\singlespace  {\small
\begin{verbatim}
LoadVectorOfVectorsOfItemsFromBank0(n, m, ass);

par{
ProduceVectorOfVectorsOfItems(bss, m, k, bssTemp);
MatrixMult(bss, css, m);
StoreVectorOfVectorsOfItems(css, n, k, c ssTemp);}
\end{verbatim}}}

%%%%%%%%%%%%%%%%%%%%%%%%%%%%%%%%%%%%%%%%%%%%%%%%%%%%%%%%%%%%%%%%%%%%%%%%%%%%%%%%%

\subsection{Second Design - Streaming I/O}

The code implementation of the current design reflects the change
as applied to the first design in the refinement to \textit{CSP}
change. The process \textit{VMMULT} definition is recalled:

$VMMULT(ass)= PRD(ass) \rhd VMMULT$

The only modification to the implementation is done by refining \textit{ass} to a stream of
vectors produced internally within \textit{VectMatrixMult}. To meet the change, this macro employs
the macro \textit{SMap} the sequential implementation of the process \textit{SMAP}. Accordingly,
the \textit{Handel-C} code is:

{\singlespace  {\small
\begin{verbatim}
 macro proc VectMatrixMult (bss, cs) {
    StreamOfVectorsOfItems(ass, n, m, Int);
    SinkVectorOfVectorsOfItems(bss, n, m, sink);
    par {
    ProduceStreamOfVectorsOfItems(Ass, n, m, tempAss);
    SMap(ass, cs, n, VScalarP);}}
\end{verbatim}}}

%%%%%%%%%%%%%%%%%%%%%%%%%%%%%%%%%%%%%%%%%%%%%%%%%%%%%%%%%%%%%%%%%%%%%%%%%%%%%%%%%

\subsection{Third Design - Pipelining}

For this design, the decomposed process for a single pipe stage is
implemented as the macro \textit{PipeStage}. This macro starts by
inputting from the left a vector \textit{bs} and a stream
\textit{cs}. These two inputs are the output of the left identical
pipe stage. For the initial pipe stage one input is the produced
stream of vectors \textit{bss} and nothing on the stream channel.
The \textit{PipeStage} then computes for the scalar product,
forward the new stream \textit{cs} and the vector \textit{bs} to
the right identical process and finally waits for another input
signal. When the end of transmission is reached, an \textit{EOT}
signal is sent to the right pipe stage. The output \textit{bss}
from the final stage is sinked, while the stream of streams output
\textit{css} is stored as the result. The code of a single pipe
stage is as follows:

{\singlespace  {\small
\begin{verbatim}
 macro proc PipeStage(tupleIn, tupleOut, iAs){
    .
    .
    .
    Item(outputItem, Int16);
    VectorOfItems(tempBs, m, Int16);
    VectorOfItems(tempAs, m, Int16);

    do{
        prialt{
        case tupleIn.element1.elements[0].channel ? tempVbs[0]:
            par(j = 1; j < m; j++){
             sOVIn.elements[j].channel ? tempVbs[j];}
             StoreStreamOfItems(tupleIn.element2.elements, iAs, tCs);
            par{
             ProduceVectorOfItems(tempBs, m, tempVbs);
             ProduceVectorOfItems(tempAs, m, ass[iAs]);
             VScalarP(tempAs, tempBs, m, outputItem);
             StoreItem(outputItem, tCs[iAs]);}

            ProduceVectorOfItems(tupleOut.element1, m, tempVbs);
            ProduceStreamOfItems(tupleOut.element2, (iAs + 1), tCs);
            break;

        case TupleIn.element1.eotChannel ? eot:
            TupleIn.element2.eotChannel ? eot1;
            TupleOut.element1.eotChannel ! True;
            TupleIn.element2.eotChannel ! True;
            break;}} while (!eot);}
\end{verbatim}}}

The general replicating macro that corresponds to the employed decomposition is implemented as in
the following code section. This macro takes the advantage of using the \textit{ifselect} and
\textit{par} \textit{Handel-C} constructs. By using \textit{ifselect}, whole statements can be
selected or discarded at compile time, depending on the evaluation of the expression. Accordingly,
the par statement selects only one macro execution for \textit{P} according to the value of
\textit{c} for each replication. The parameter \textit{c} corresponds to the pipe stage number. In
this design this parameter is initialised to \textit{1} instead of \textit{0} since a different
initial pipe stage is implemented to overcome a limitation in \textit{Handel-C}. This limitation
forbids the production of stream of streams with a size \textit{0} needed for the first pipe
stage. The final picture of this implementation is best depicted as in Figure~\ref{fig9}.

{\singlespace  {\small
\begin{verbatim}
 macro proc Pipe (tIn, tOut, n, P){
    typeof(tIn) cmids[n - 1];

        par (c = 1; c < n; c++){
            ifselect (c == 1)
               P(tIn, cmids[c], c);
            else ifselect (c < n - 1)
                P(cmids[c - 1], cmids[c], c);
            else
                P(cmids[c - 1], TOut, c);}}
\end{verbatim}}}

The execution of the matrix multiplication is done as follows.

{\singlespace  {\small
\begin{verbatim}
 par {
 ProduceStreamOfVectorsOfItemsFromBank0(bssIn, m, bss);
 PipeStageInitial(bssIn, tIn, 0);
 Pipe (tIn, tOut, n, PipeStage);
 StoreStreamOfStreamsOfItemsInBank1(tOut.element2, n, tempOut1);
 SinkStreamOfVectorsOfItems(tOut.element1, m, tempOut2);}
\end{verbatim}}}

%%%%%%%%%%%%%%%%%%%%%%%%%%%%%%%%%%%%%%%%%%%%%%%%%%%%%%%%%%%%%%%%%%%%%%%%%%%%%%%%%

\subsection{Fourth Design - Pipelined Turnout Stages}

In a similar way of implementing the third design, this design uses a modified version of the
previous macros. In the new \textit{TurnoutPipeStage} the output is a vector of streams
\textit{vOSOUT}. Each pipe stage outputs its own stream and issues its own termination signal
through its own \textit{eot} channel. This macro works by firstly inputting a vector from the
stream \textit{bss}, forwards it to the right stage, then computes for the scalar product turning
out its result. These steps are repeated till the end of transmission of \textit{bss}. At that
point, local termination signals are issued from the stage. The final picture of this
implementation is best depicted as in Figure~\ref{fig10}.

{\singlespace  {\small
\begin{verbatim}
 macro proc TurnoutPipeStage(sOVIn, sOVOut, vOSOut, indexForAs) {
    .
    .
    .
    VectorOfItems(tempBs, m, Int16);
    VectorOfItems(tempAs, m, Int16);
    do{
    prialt{
        case sOVIn.elements[0].channel ? tempVbs[0]:
            par(j = 1; j < m; j++){
                sOVIn.elements[j].channel ? tempVbs[j];}
            ProduceVectorOfItems(sOVOut, m, tempVbs);
            par{
                ProduceVectorOfItems(tempBs, m, tempVbs);
                ProduceVectorOfItems(tempAs, m, ass[indexForAs]);
                VScalarP(tempAs, tempBs, m, vOSOut);}
            break;
        case sOVIn.eotChannel ? eot:
            sOVOut.eotChannel ! True;
            vOSOut.eotChannel ! True;
            break;}} while (!EOT);}
\end{verbatim}}}

This pipe pattern is implemented as:

{\singlespace  {\small
\begin{verbatim}
 macro proc TurnoutPipe(in1, out1, out2, n, p) {
 typeof(in1) cmids1[n + 1];

    par (c = 0; c < n; c++){
        ifselect (c == 0)
            p (in1, cmids1[c], out2.elements[c], c);
        else ifselect (c < n - 1)
            p (cmids1[c - 1], cmids1[c], out2.elements[c], c);
        else
            p(cmids1[c - 1], out1, out2.elements[c], c);}}
\end{verbatim}}}

The execution of this design's implementation is done as follows:

{\singlespace  {\small
\begin{verbatim}
 void main(void) {
    .
    .
    .
    StreamOfVectorsOfItems(bssIn, m, Int16);
    StreamOfVectorsOfItems(bssOut, m, Int16);
    VectorOfStreamsOfItems(cssOut, n, Int16);
    par {
            ProduceStreamOfVectorsOfItems(bssIn, n, m, ass);
            TurnOutPipe(bssIn, bssOut, cssOut, 3, TurnoutPipeStage);
            StoreVectorOfStreamsOfItems(cssOut, n, k, tempCss);
            SinkStreamOfVectorsOfItems(bssOut, m, tempBss);}
    StoreMatrixInBank2(tempBss, n, k);}
\end{verbatim}}}

It is clear from this design that n-parallel output streams are employed. Consequently, a buffer
with multi-concurrent access is needed. This kind of access is not allowed with the available
single \textit{R/W} onboard \textit{SRAMs}. Besides, the number of available banks for concurrent
access is only 4. This introduces a limitation to the practical implementation of this design:

\begin{itemize}
\item The parameter \textit{k} will appear as a
static constant in the compilation. Thus, for any new stream with a certain length a new
compilation is needed.

\item The limited
ability of the \textit{FPGA} to store the results on its internal area, especially, for matrices
with large dimensions.
\end{itemize}

the suggested general solution to this problem is storing the resultant
\textit{VectorOfStreamsOfItems} on the local \textit{FPGA} memory, and then storing them back as a
stream of values or a vector of streams with up to 4 parallel streams only.

%%%%%%%%%%%%%%%%%%%%%%%%%%%%%%%%%%%%%%%%%%%%%%%%%%%%%%%%%%%%%%%%%%%%%%%%%%%%%%%%%

\subsection{Multilevel Pipelines Design}

This design implementation uses two kinds of pipe replicating
macros for the two needed pipelined levels. The first pipe
implements the decomposition for the vector matrix multiplication
process. Thus, we reuse the macro \textit{TurnoutPipe} employed in
the fourth design. Furthermore, the second pipe macro, called
\textit{SystolicPipe}, implements the scalar product process. This
macro replicates pipe stages with two inputs and two outputs.
Recall the \textit{CSP} implementation for a the basic computation
cell:

$CELL (a)= PRD(a) \rhd (up ? u \rightarrow left? l  \rightarrow right!(u*a + l) \rightarrow down!u
)$

The code corresponding to \textit{CELL} is:

{\singlespace  {\small
\begin{verbatim}
 macro proc SystolicPipeCell(li, upV, r, downV, i, j) {
    .
    .
    .
    Item(temp, Int16);

    upV.Channel ? tempb;
    result = tempb * ass[i][j];
    par{
        Addition(temp, li, r);
        ProduceItem(temp, result);}
    downV.Channel ! tempb;}
\end{verbatim}}}

To implement the complete algorithm, these cells are composed to form a scalar product pipeline
using the macro \textit{SystolicPipe}.

{\singlespace  {\small
\begin{verbatim}
 macro proc SystolicPipe (in1l, in2up, out1r, out2d, n, i,P){
 typeof(in1l) cmids1[n + 1];

 par (c = 0; c < n; c++){
 ifselect (c == 0)
    P(in1l,in2up.elements[c],cmids1[c],out2d.elements[c],i,c);
 else ifselect (c < n - 1)
    P(cmids1[c-1],in2up.elements[c],cmids1[c],out2d.elements[c],i,c);
 else
    P(cmids1[c-1],in2up.elements[c],out1r,out2d.elements[c],i,c);}}
\end{verbatim}}}

This scalar product pipe is the core of the pipe stage needed for the pipeline implementing the
matrix multiplication algorithm. A pipe stage is implemented as:

{\singlespace  {\small
\begin{verbatim}
 macro proc PipeStage(bssIn, bssOut, cssOut, i){
 VectorOfItems(tempBs, m, Int16);
 Item(rO, Int16);
 Item(lI, Int16);
 .
 .
 .
 do{
    prialt{
        case bssIn.elements[0].channel ? temp[0]:
          par(j = 1; j < m; j++){
            bssIn.elements[j].channel ? temp[j];}
          par{
            ProduceItem(lI, 0);
            ProduceVectorOfItems(tempBs, m, temp);
            SystolicPipe(lI, tempBs, rO, cssOut, m, i, SystolicPipeCell);
            StoreItem(rO, tempItem);}
          break;

        case bssIn.eotChannel ? eot:
          bssOut.eotChannel ! eot;
          break;}} while (!eot);}
\end{verbatim}}}

This pipe stage is then replicated to form a pipeline using the predefined macro
\textit{TurnoutPipe}. The main code section running the above implementation is:

{\singlespace  {\small
\begin{verbatim}
 void main(void) {
    .
    .
    .
    StreamOfVectorsOfItems(bssIn, m, Int16);
    StreamOfVectorsOfItems(bssOut, m, Int16);
    StreamOfVectorsOfItems(cssFinal, n, Int16);

    par{
        ProduceStreamOfVectorsOfItemsFromBank1(bssIn, m);
        {TurnoutPipe(bssIn, bssOut, cssFinal, n, PipeStage);
        cssFinal.eotChannel !True;}
        SinkStreamOfVectorsOfItemsToBank3(bssOut, m);
        StoreStreamOfVectorsOfItemsInBank2(cssFinal,  n);}}
\end{verbatim}}}

%%%%%%%%%%%%%%%%%%%%%%%%%%%%%%%%%%%%%%%%%%%%%%%%%%%%%%%%%%%%%%%%%%%%%%%%%%%%%%%%%

\section{Performance Analysis and Evaluation} \label{PAE}

The development is originated from a specification stage, whose
main key feature is its powerful \textbf{higher-level of
abstraction}. During the specification, the isolation from
parallel hardware implementation technicalities allowed for deep
concentration on the specification details. Whereby, for the most
part, the style of specification comes out in favor of using
higher-order functions. Two other inherent advantages for using
the functional paradigm are \textbf{clarity} and
\textbf{conciseness} of the specification. This was reflected
throughout all the presented studies. At this level of
development, the \textbf{correctness} of the specification is
insured by construction from the used correct building blocks. The
implementation of the formalised specification is tested under
\textit{Haskell} by performing random tests for every level of the
specification.

The correctness will be carried forward to the next stage of development by applying the provably
correct rules of refinement. The available pool of refinement formal rules enables a high degree
of \textbf{flexibility} in creating parallel designs. This includes the capacity to divide a
problem into completely independent parts that can be executed simultaneously (pleasantly
parallel). Conversely, in a nearly pleasantly parallel manner, the computations might require
results to be distributed, collected and combined in some way. Remember at this point, that the
refinement steps are \textbf{systematic} and done by combining off-the-shelf \textbf{reusable}
instances of basic building blocks.

In this case study, we will measure the speed in Items per Second (ips). For instance, a
$3\times3$ matrix has 9 items.

In Table~\ref{TblResults} the results for running the different designs are presented. The first
design occupied an area of 564 Slices per Item running at a speed of 257.14 Kips for a network
with dimensions of $(3\times3\times3)$. The second design occupied a smaller area, as expected,
but the realised design runs with a speed of 280 Kips. The pipelined second and third designs
achieved a better speed of 2.25 Mips with less areas of 227 and 237 Slices per Item. The smallest
area ratio of 158.7 Slices per Item has been occupied when placing and routing the 2D pipelined
design. The speed achieved by this design is 3.1 Mips.

\begin{landscape}
\begin{table} [htpb]
	\caption{The results of testing the suggested matrix multiplication designs}
	\label{TblResults}
	\begin{center}
		\includegraphics [scale= 1]%[height=6in,width=5in,angle=0]
		{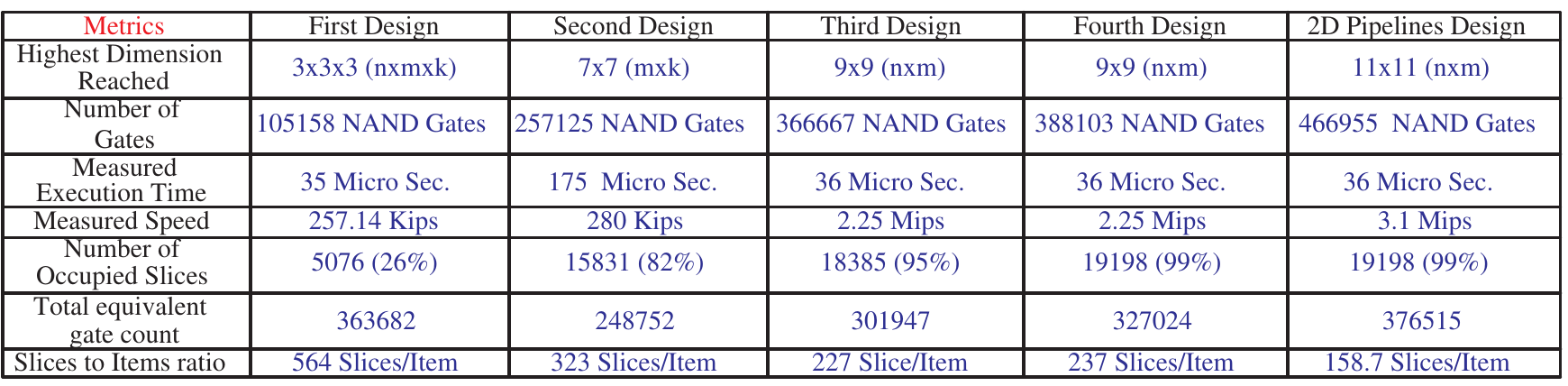}
	\end{center}
\end{table}
\end{landscape}

The second design is found to be $8.9\%$ faster than the first design, also the Slices to Items
ratio of the second design is 42.7$\%$ less than that of the first design. Thus, the second design
can accommodate for a larger number of items with a better speed as compared to the first design.
The modification done to the third pipelined design yielding the fourth pipelined design shows
that there were no effect on the speed of execution. However, the fourth design occupied a 4.4$\%$
larger area. Thus, the modification didn't leave a positive effect on the performance. The 2D
pipelined design has shown a better performance than the other designs, for instance, it occupies
a 30.1$\%$ less Slices per Item than the third pipelined design, also achieving a 38$\%$ higher
speed.

\begin{figure} [htpb]
	\begin{center}
		\includegraphics [scale = 0.85]
		{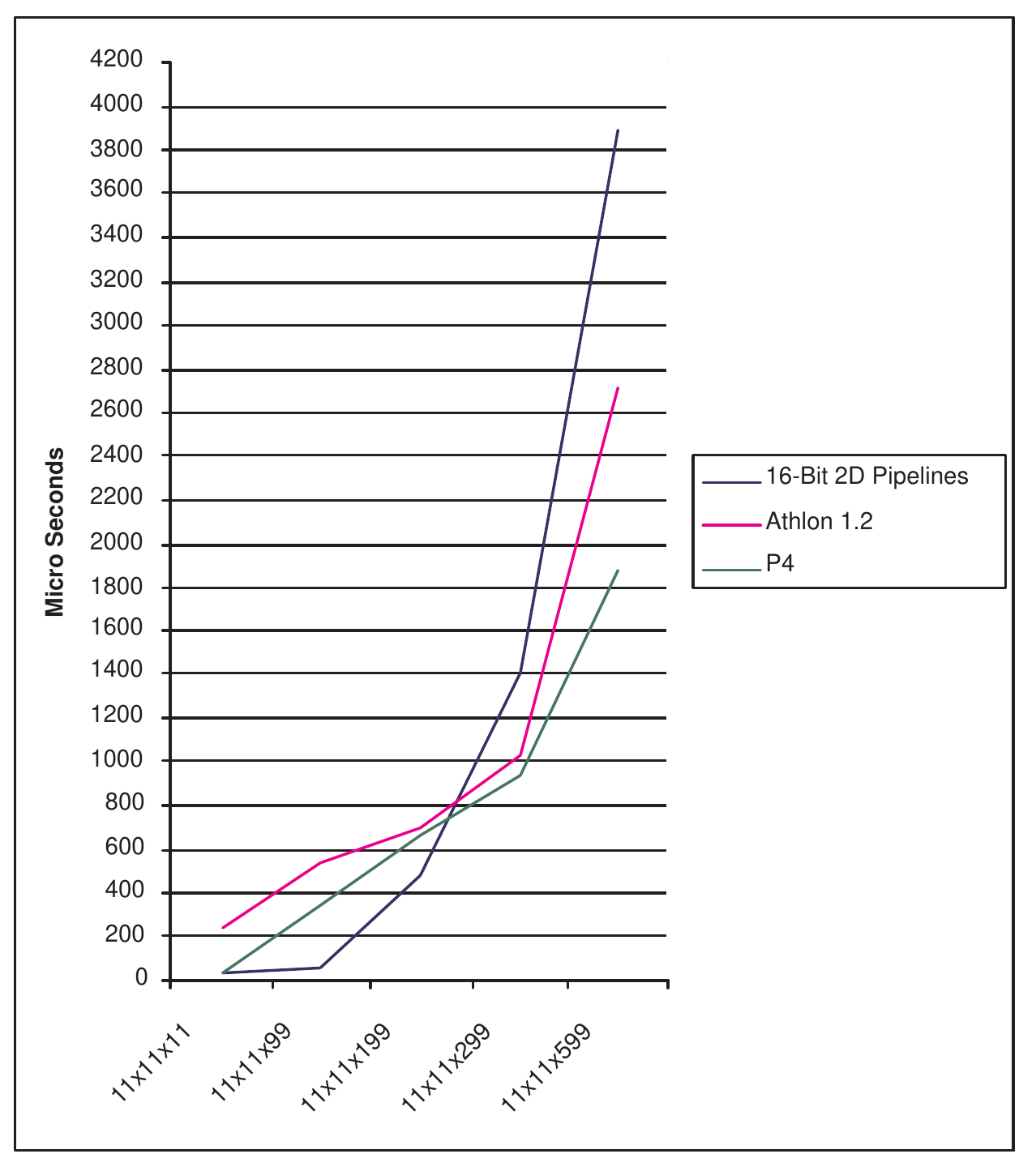}
		\caption{A chart showing the change in execution wrt dimension as shown in Table~\ref{TblCompareP} for small values of k}
		\label{Chart2}
	\end{center}
\end{figure}

\begin{figure} [htpb]
	\begin{center}
		\includegraphics [scale = 0.85]
		{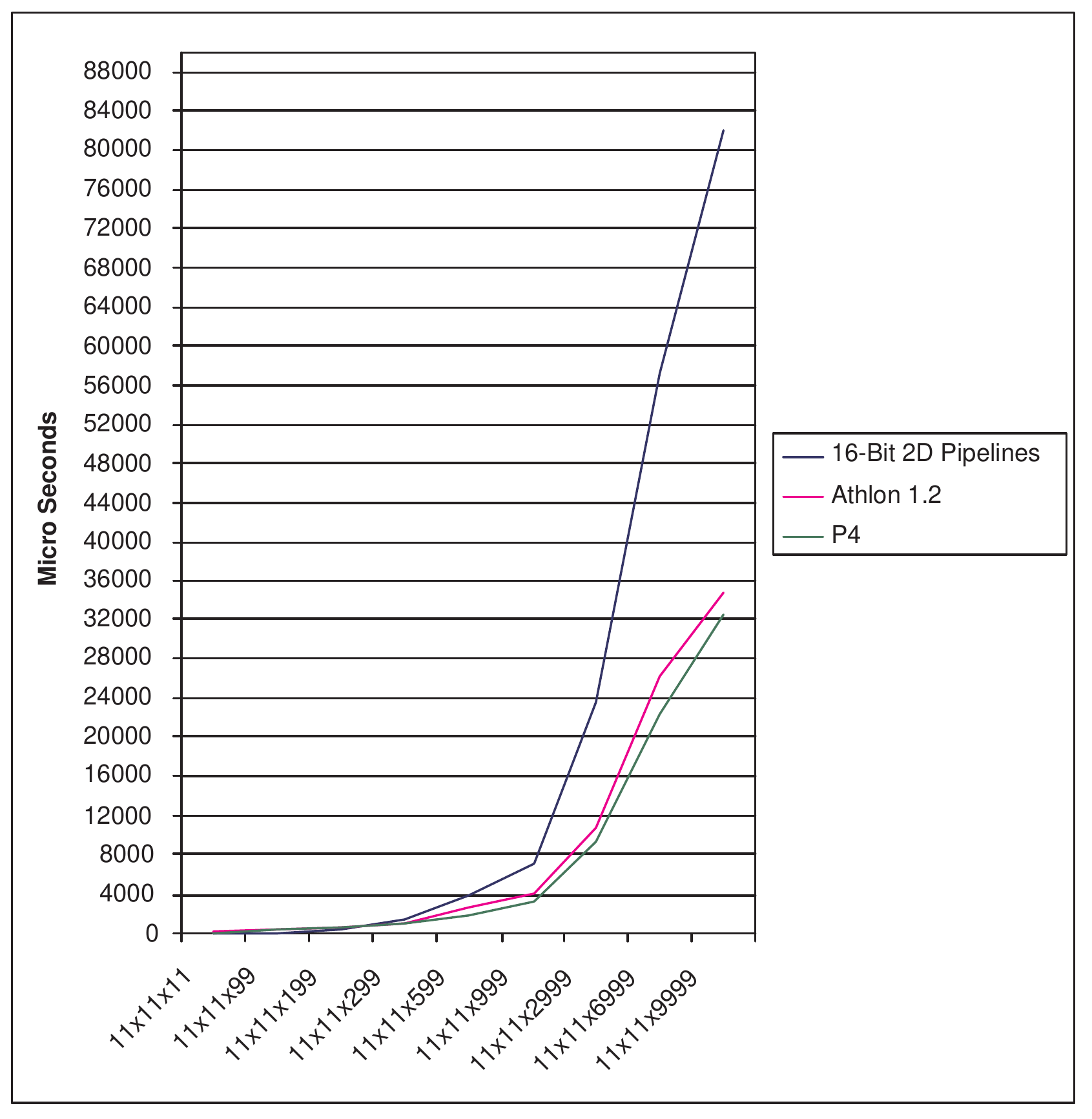}
		\caption{A chart showing the change in execution wrt dimension as shown in Table~\ref{TblCompareP}}
		\label{Chart}
	\end{center}
\end{figure}

The $(11 \times 11)$ 2D pipelined cells design is independent from the third dimension $k$. In
Table~\ref{TblCompareP} we compare the execution time of running for different values of $k$
between the \textit{RC-1000} and two computer machines. These are a 1.2 GHz \textit{Athlon}
\textit{AMD} machine with 512MB of \textit{RAM}, and a 1.4 GHz \textit{P4} with 1GB of
\textit{RAM}. It is shown in Figures~\ref{Chart} and ~\ref{Chart2} that these machines will
outperform the suggested design implementation on the \textit{RC-1000} when the value of $k$ is
nearly at a value of $299$ items. We note here the possible effect of the bus connecting the
memory and \textit{FPGA} on the speed of execution. To cope with this limitation a suggestion
could be proposed for adding a cache memory to handle the input and output streams of data.

\begin{table} [htpb]
	\caption{Comparisons between the results of testing the 2D pipelined design and a C++ implementation running on two
		different personal computing machines; the results shown are in Micro Seconds}
	\label{TblCompareP}
	\begin{center}
		\includegraphics [scale= 1]%[height=6in,width=5in,angle=0]
		{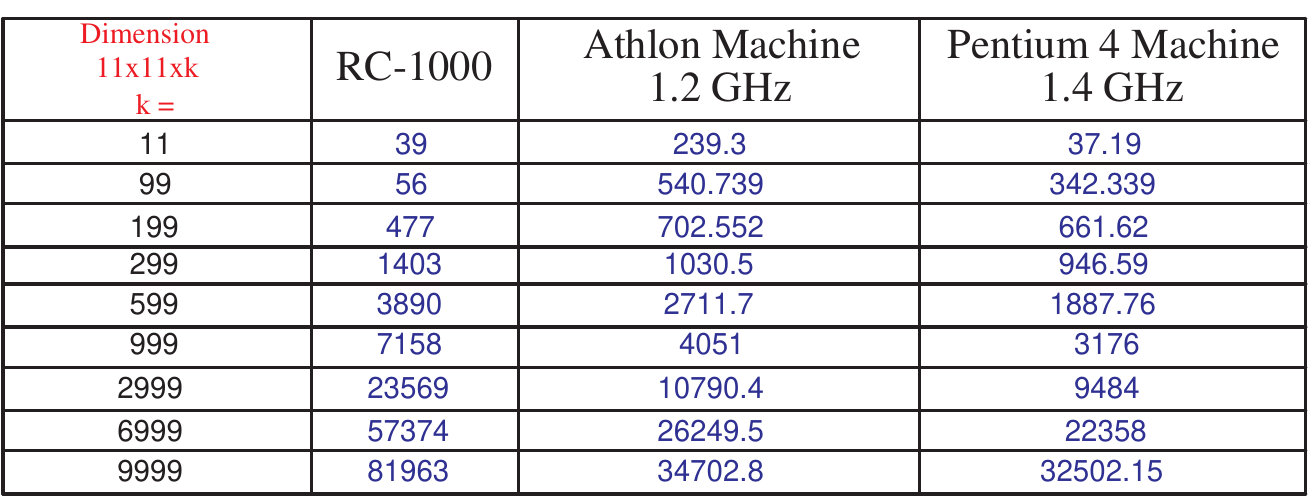}
	\end{center}
\end{table}

%%%%%%%%%%%%%%%%%%%%%%%%%%%%%%%%%%%%%%%%%%%%%%%%%%%%%%%%%%%%%%%%%%%%%%%%%%%%%%%%%

\section{Acknowledgement}
I would like to thank Dr. Ali Abdallah, Prof. Mark Josephs, Prof. Wayne Luk, Dr. Sylvia Jennings,
and Dr. John Hawkins for their insightful comments on the research which is partly presented in
this paper.

%%%%%%%%%%%%%%%%%%%%%%%%%%%%%%%%%%%%%%%%%%%%%%%%%%%%%%%%%%%%%%%%%%%%%%%%%%%%%%%%%

\section{Conclusion} \label{Con}
Mapping parallel versions of algorithms onto hardware could enormously improve computational
efficiency. Recent advances in the area of reconfigurable computing came in the form of
\textit{FPGAs} and their high-level \textit{HDLs} such as \textit{Handel-C}. In this paper, we
build on these recent technological advances by presenting, demonstrating and examining a
systematic approach of behavioural synthesis. This system creates a functional specification of an
algorithm without defining parallelism. Correspondingly, an efficient parallel implementation is
derived in the form of \textit{CSP} network of processes. Accordingly, we create efficient
parallel implementations in \textit{Handel-C}. The presented work included theory and practices
about the suggested methodology. This paper also presented a demonstration for using a proposed
model to synthesise reconfigurable hardware for the matrix multiplication algorithm. The general
functional specification is discussed firstly followed by the provably correct step-wise refitment
to \textit{CSP}. Many possible designs were engineered and compiled to reconfigurable hardware
with different levels of parallelism. The hardware implementation using \textit{Handel-C} is shown
stressing the correspondence to the \textit{CSP} refined networks. To complete the synthesis,
these designs were compiled to \textit{EDIF} format and then placed and routed. Accordingly, a
performance study has been included for the realised designs. The first design required the
largest area with respect to the number of items used, that is 564 Slices per Item for a speed of
257.14 Kips. The modification to the first design which lead to the second design helped in
reducing the area to 323 Slices per Item for a speed of 280 Kips. The third and fourth pipelined
designs occupied areas of 227 and 237 Slices per Item for a speed of 2.25 Mips. The lastly
realised 2D pipelines design has the best area to items ratio of 158.7 Slices per Item and a speed
of 3.1 Mips. Future work includes extending the theoretical pool of rules for refinement, the
investigation of automating the development processes, and the optimisation of the realisation for
more economical implementations with higher throughput.

%%%%%%%%%%%%%%%%%%%%%%%%%%%%%%%%%%%%%%%%%%%%%%%%%%%%%%%%%%%%%%%%%%%%%%%%%%%%%%%%%
\newpage

{\singlespace

\bibliography{MM}

\begin{thebibliography}{10}
\expandafter\ifx\csname url\endcsname\relax
  \def\url#1{\texttt{#1}}\fi
\expandafter\ifx\csname urlprefix\endcsname\relax\def\urlprefix{URL }\fi

\bibitem{59}
G.~Estrin, B.~Bussell, R.~Turn, J.~Bibb, Parallel processing in a
  restructurable computer system, IEEE Transactions on Electronic Computers
  12~(6) (1963) 747--755.

\bibitem{126}
Xilinx, Information available from, http://www.xilinx.com.

\bibitem{127}
Altera, Information available from, http://www.Altera.com.

\bibitem{128}
Altium, Altium unveils new "board-on-chip" technology, Altium Limited Category:
  Press Releases : Industry News (Market) http://www.altium.com (April 2003).

\bibitem{4}
K.~Torkelsson, J.~Ditmar, {Header Compression in Handel-C An Internet
  Application and A New Design Language}, in: Symposium on Digital Systems
  Design, Euromicro, 2001, pp. 2--7.

\bibitem{129}
Celoxica, Information available from, http://www.celoxica.com.

\bibitem{55}
I.~Page, Logarithmic greatest common divisor example in {Handel-C}, Embedded
  Solutions (April 1998).

\bibitem{125}
S.~Stepney, {CSP/FDR2} to {Handel-C} translation, Tech. Rep. YCS-2002-357,
  Department of Computer Science, University of York (June 2003).

\bibitem{130}
D.~Edwards, S.~Harris, J.~Forge, High performance hardware from java, Xilinx
  Whitepaper http://www.xilinx.com.

\bibitem{131}
Y.~Li, T.~Callahan, E.~Darnell, R.~Harr, U.~Kurkure, J.~Stockwood,
  Hardware-software codesign of embedded reconfigurable architectures, in:
  Proceedings of the 37th Design Automation Conference, Los Angeles - USA,
  2000, p.~30.

\bibitem{132}
N.~Technology, Information available from, http://www.nimble.com.

\bibitem{133}
S.~Network, Information available arom, http://www.systemc.org.

\bibitem{newrev1}
Viva, Information available from, http://www.starbridgesystems.com.

\bibitem{6}
A.~E. Abdallah, Functional process modelling, Research Directions in Parallel
  Functional Programming, (Springer Verlag, October 1999) (1999) 339--360.

\bibitem{9}
A.~E. Abdallah, {D}erivation of {P}arallel {A}lgorithms: {F}rom {F}unctional
  {S}pecifications to csp {P}rocesses, in: B.~Moller (Ed.), Proceedings of
  Mathematics of Program Construction, Vol. 947 of Lecture Notes in Computer
  Science, Springer-Verlag, 1994, pp. 67--96.

\bibitem{140}
A.~E. Abdallah, J.~Hawkins, {C}alculational {D}esign of {S}pecial {P}urpose
  {P}arallel {A}lgorithms, in: Proceedings of 7th IEEE International Conference
  on Electronics, Circuits and Systems (IEEE/ICECS), IEEE Computer Society
  Press, 2000, pp. 261--267.

\bibitem{140b}
A.~E. Abdallah, J.~Hawkins, {F}ormal {B}ehavioural {S}ynthesis of handel-c
  {P}arallel {H}ardware {I}mplementation for {F}unctional {S}pecifications, in:
  Proceedings of the 36th Annual Hawaii International Conference on System
  Sciences, IEEE Computer Society Press, 2003, pp. 278--288.

\bibitem{B4}
Y.~Lee, B.~Ryder, A comprehensive approach to parallel data flow analysis, in:
  Proceedings of the 6th international conference on Supercomputing, ACM Press,
  1992, pp. 236--247.

\bibitem{10}
R.~Bird, P.~Wadler, Introduction to Functional Programming, Prentice-Hall,
  1988.

\bibitem{11}
R.~Bird, Introduction to Functional Programming Using Haskell, Addison Wesley,
  1999.

\bibitem{Brd87}
R.~Bird, An introduction to the theory of lists, in: M.~Broy (Ed.), Logic of
  Programming and Calculi of Discrete Design, Springer, Berlin, Heidelberg,
  1987, pp. 5--42.

\bibitem{13}
M.~Cole, {Algorithmic Skeletons: A Structured Approach to the Management of
  Parallel Computation}, Ph.D. thesis, Computer Science Department, University
  of Edinburgh, Edinburgh, Scotland, UK (1988).

\bibitem{14}
J.~Darlington, A.~Field, P.~Harrison, H.~Paul, J.~Kelly, D.~Sharp, Q.~Wu,
  Parallel programming using skeleton functions, in: Proceedings of the 5th
  International PARLE Conference on Parallel Architectures and Languages
  Europe, Springer-Verlag, 1993, pp. 146--160.

\bibitem{15}
S.~Gorlatch, C.~Lengauer, Parallelization of divide-and-conquer in the
  bird-meertens formalism, Formal Aspects of Computing 7~(6) (1995) 663--682.

\bibitem{Rbh95}
F.~A. Rabhi, {Exploiting Parallelism in Functional Languages: A
  ''Paradigm-Oriented`` Approach}, in: T.~Lake, P.~Dew (Eds.), Abstract Machine
  Models for Highly Parallel Computers, Oxford University Press, 1993, p.~30.

\bibitem{RnW95}
F.~Hanna, W.~Howells, Parallel {T}heorem {P}roving, in: C.~Runciman,
  D.~Wakeling (Eds.), Applications of Functional Programming, UCL Press, 1994,
  Ch.~12, pp. 221-- 235.
\newline\urlprefix\url{http://www.cs.ukc.ac.uk/pubs/1994/432}

\bibitem{Skl94}
D.~Skillicorn, Foundations of Parallel Programming, Cambridge University Press,
  1994.

\bibitem{122}
K.~Claessen, Embedded languages for describing and verifying hardware, Ph.D.
  thesis, Chalmers Univesity of Technology and G$\ddot{o}$teborg University,
  Sweden (April 2001).

\bibitem{r1}
J.~Launchbury, J.~Lewis, B.~Cook, On embedding a microarchitectural design
  language within haskell, in: Proceedings of the fourth ACM SIGPLAN
  international conference on Functional programming, ACM Press, 1999, pp.
  60--69.

\bibitem{r2}
J.~Matthews, J.~Launchbury, B.~Cook, Specifying microprocessors in hawk, in:
  Proceedings of the International Conference on Computer Languages, IEEE,
  1998, pp. 90--101.

\bibitem{r4}
J.~O'Donnell, {Hydra}: hardware description in a functional language using
  recursion equations and high order combining forms, in: G.~J. Milne (Ed.),
  The Fusion of Hardware Design and Verification, North-Holland, Amsterdam,
  1988, pp. 309--328.

\bibitem{r7}
Y.~Li, M.~Leeser, {HML}: An innovative hardware design language and its
  translation to {VHDL}, in: Conference on Hardware Design Languages, 1995.

\bibitem{r6}
D.~Barton, Advanced modeling features of {MHDL}, in: In International
  Conference on Electronic Hardware Description Languages, 1995.

\bibitem{r8}
S.~Johnson, B.~Bose, {DDD}: A system for mechanized digital design derivation,
  Tech. Rep. 323, Indiana University, Indiana (1990).

\bibitem{r10}
R.~Sharp, Higher-level hardware synthesis, Ph.D. thesis, Robinson College
  University of Cambridge, Cambridge (November 2002).

\bibitem{r11}
M.~Sheeran, mu{FP}: a language for {VLSI} design, in: Proc. ACM Symposium on
  LISP and Functional Programming, ACM Press, 1984, pp. 104--112.

\bibitem{r12}
G.~Jones, M.~Sheeran, Circuit design in ruby, In Formal Methods for VLSI design
  (1990) 13--70.

\bibitem{r14}
T.~Cheung, G.~Hellestrand, Multi-level equivalence in design transformation,
  in: Proceedings of International Conference on Computer Hardware Description
  Languages, Chiba Japan, 1996, pp. 559--566.

\bibitem{16}
C.~A.~R. Hoare, Communicating Sequential Processes, Prentice-Hall, 1985.

\bibitem{8}
A.~E. Abdallah, Synthesis of massively pipelined algorithms for list
  manipulation, in: L.~Bouge, P.~Fraigniaud, A.~Mignotte, Y.~Robert (Eds.),
  Proceedings of the European Conference on Parallel Processing, EuroPar'96,
  LNCS 1024, (Springer Verlag, 1996), Springer Verlag, 1996, pp. 911--920.

\bibitem{B3}
I.~Ltd., {OCCAM} 2 reference manual, Prentice-Hall International (1988).

\bibitem{MM6}
E.~Horowitz, A.~Zorat, Divide-and-conquer for parallel processing, IEEE Trans.
  Comput. C32~(6) (1983) 582--585.

\bibitem{MM7}
J.~Hake, Parallel algorithms for matrix operations and their performance on
  multiprocessor systems, Advances in Parallel Algorithms.

\bibitem{MM8}
G.~Fox, M.~Johnson, G.~Lyenga, S.~Otto, J.~Salmon, D.~Walker, Solving Problems
  on Concurrent Processors, Vol.~1, Prentice Hall, Emglewood Cliffs, New Jersy,
  1988.

\bibitem{MM5}
L.~Cannon, A celluler computer to implement the kalman filter algorithm, {Ph.D.
  Thesis}, Montana State University, Bozman - Montana (1969).

\bibitem{MM1}
J.~Choi, J.~Dongarra, R.~Pozo, D.~Walker, {SCALAPACK}: A scalable linear
  algebra library for distributed memory concurrent computers, in: Proceedings
  of the Fourth Symposium on the Frontiers of Massively Parallel Computation,
  IEEE Comput. Soc., 1992, pp. 120--127.

\bibitem{MM2}
J.~Choi, J.~Dongarra, D.~Walker, {PUMMA}: Parallel universal matrix
  multiplication algorithms on distributed memory concurrent computers,
  Concurrency: Practice and Experience 6 (1994) 543--57.

\bibitem{MM3}
J.~Dongarra, I.~Duff, D.~Sorensen, H.~V.~D. Vorst, Solving linear systems on
  vector and shared memory computers, in: SIAM, 1991, p.~30.

\bibitem{MM4}
S.~Huss-Lederman, E.~Jacobson, A.~Tsao, Comparison of scalable parallel matrix
  multiplication libraries, in: Proceedings of the Scalable Parallel Libraries
  Conference, IEEE Comput. Soc., Starksville - MS, 1993, pp. 120--127.

\end{thebibliography}
\bibliographystyle{elsart-num}

\newpage

%%%%%%%%%%%%%%%%%%%%%%%%%%%%%%%%%%%%%%%%%%%%%%%%%%%%%%%%%%%%%%%%%%%%%%%%%%%%%%%%%

\section{Biography for Issam Damaj}
Issam W. Damaj (Ph.D. M.Eng. B.Eng. MIEEE MIEE) received his B.Eng. in Computer Engineering from
Beirut Arab University in 1999 (with high distinction), and his M.Eng. in Computer and
Communications Engineering from the American University of Beirut in 2001 (with high distinction).
He was awarded his Ph.D. degree in Computer Science from London South Bank University, London,
United Kingdom in 2004. Currently, he is with the Electrical and Computer Engineering Department
at Hariri Canadian Academy for Sciences and Technology, Lebanon. His research interests include
reconfigurable computing, parallel processing, h.w./s.w. co-design, computer interfacing and
applications, fuzzy logic, and computer security. He has more than 25 international and regional
research publications and projects. He is a Member of the IEEE and IEE professional organizations,
and the order of Engineers in Beirut.

%
%\newpage
%
%\listoffigures
%
%\newpage
%
%\listoftables
%
%\newpage

%%%%%%%%%%%%%%%%%%%%%%%%%%%%%%%%%%%%%%%%%%%%%%%%%%%%%%%%%%%%%%%%%%%%%%%%%%%%%%%%%

}%end single space
%%%%%%%%%%%%%%%%%%%%%%%%%%%%%%%%%%%%%%%%%%%%%%%%%%%%%%%%%%%%%%%%%%%%%%%%%%%%%%%%%
\end{document}